\documentclass[12pt,titlepage]{article}
\usepackage[utf8]{inputenc}
\usepackage[margin=1in,letterpaper]{geometry}
\usepackage[nodisplayskipstretch]{setspace}
\usepackage{mathtools,amssymb,microtype,bm,dsfont,graphicx,color,soul,placeins,booktabs,multicol,multirow,rotating,natbib}
\usepackage{adjustbox,enumitem,wasysym,titlesec,array}
\usepackage{tabularx}
\usepackage{pdflscape}

\usepackage{titling}
\pretitle{\vspace*{\fill}\begin{center}\Large}
\posttitle{\par\end{center}\vskip 2em}
\preauthor{\begin{center}\normalsize\lineskip 1em \begin{tabular}[t]{c}}
\postauthor{\end{tabular}\par\end{center}}
\predate{\begin{center}\normalsize\vskip 2em}
\postdate{\end{center}\vskip 0em
  \begin{center}\ifdefined\draftnote\emph{\draftnote}\fi\end{center}
  \vspace*{\fill}
  \ifdefined\acknowledgments\ifx\acknowledgments\empty\else
  \begin{singlespace}\footnotesize\noindent\emph{Acknowledgments:} \acknowledgments\end{singlespace}
  \fi\fi
}
\let\origmaketitle\maketitle
\renewcommand{\maketitle}{\begingroup\onehalfspacing\origmaketitle\endgroup}

\AtBeginEnvironment{abstract}{\onehalfspacing}
\AtBeginEnvironment{thebibliography}{\singlespacing}

\bibpunct[, ]{(}{)}{,}{a}{}{,}

\setcitestyle{notesep={:}}
\newcommand{\p}[1]{#1}

\titleformat{\subsubsection}[runin]{\normalfont\itshape}{\thesubsubsection}{1em}{}[.\enspace]
\titlespacing{\subsubsection}{\parindent}{\parskip}{0pt}

\usepackage{listings}
\usepackage{caption}
\DeclareCaptionFormat{listing}{\noindent\normalsize\textbf{#3}}
\usepackage{dcolumn}
\newcolumntype{d}[1]{D{.}{.}{#1}}
\let\oldappendix\appendix
\renewcommand{\appendix}{\oldappendix\titleformat{\section}[block]{\normalfont\Large\bfseries}{Appendix~\thesection:~}{0em}{}\setcounter{figure}{0}\numberwithin{figure}{section}\setcounter{table}{0}\numberwithin{table}{section}}

\usepackage[breaklinks,hidelinks]{hyperref}
\usepackage{xurl}

\newcommand{\Keywords}{\bigskip\noindent\textbf{Keywords:\ }}

\author{
Felipe A. Csaszar \\
Ross School of Business \\
University of Michigan \\
Ann Arbor, MI 48109 \\
\texttt{fcsaszar@umich.edu}
\and
Aticus Peterson \\
NYU Stern School of Business \\
New York University \\
New York, NY 10012 \\
\texttt{ap9674@stern.nyu.edu}
\and
Daniel Wilde \\
Kelley School of Business \\
Indiana University \\
Bloomington, IN 47405 \\
\texttt{wilded@iu.edu}
}
\date{\today}
\newcommand{\draftnote}{Draft: Please do not circulate or cite without the permission of the authors.}

\title{The Strategic Foresight of LLMs:\\Evidence from a Fully Prospective Venture Tournament}

\begin{document}
\maketitle

\begin{abstract}
Can artificial intelligence outperform humans at strategic foresight---the capacity to form accurate judgments about uncertain, high-stakes outcomes before they unfold?  We address this question through a fully prospective prediction tournament using live Kickstarter crowdfunding projects.  Thirty U.S.-based technology ventures, launched after the training cutoffs of all models studied, were evaluated while fundraising remained in progress and outcomes were unknown.  A diverse suite of frontier and open-weight large language models (LLMs) completed 870 pairwise comparisons, producing complete rankings of predicted fundraising success.  We benchmarked these forecasts against 346 experienced managers recruited via Prolific and three MBA-trained investors working under monitored conditions.  The results are striking: human evaluators achieved rank correlations with actual outcomes between 0.04 and 0.45, while several frontier LLMs exceeded 0.60, with the best (Gemini 2.5 Pro) reaching 0.74---correctly ordering nearly four of every five venture pairs.  These differences persist across multiple performance metrics and robustness checks.  Neither wisdom-of-the-crowd ensembles nor human-AI hybrid teams outperformed the best standalone model.

\Keywords artificial intelligence; strategic foresight; strategic decision-making; opportunity evaluation; strategic uncertainty
\end{abstract}

\section{Introduction}

\subsection{Can AI Make Strategic Predictions?}

In 1943, over lunch at Bell Labs, Alan Turing and Claude Shannon debated what kind of artificial intelligence they hoped to build.  Shannon envisioned a machine of vast intellectual power---what we might now call an AI scientist.  Turing countered with a different ambition.  ``No, I'm not interested in developing a powerful brain,'' he quipped.  ``All I am after is just a mediocre brain, something like the President of the American Telephone and Telegraph Company'' \citep[\p 251]{hodges_alan_1983}.  He imagined a machine that could digest facts about commodity prices and stocks, then answer the question: ``Do I buy or sell?''  In modern terms, Turing was describing an AI CEO.

Eight decades later, the asymmetry between these two visions is stark.  AI scientists now autonomously discover physical laws, design proteins, and generate novel hypotheses \citep{jumper_highly_2021, fang_ai-newton_2025, ghafarollahi_sciagents_2025}.  Yet the AI CEO remains unrealized.  Sam Altman captures this gap in a thought experiment: ``What would have to happen for an AI CEO to be able to do a much, much better job of running OpenAI than me, which clearly will happen someday.  But how can we accelerate that?  What's in the way of that?'' \citep{altman_sam_2025}.  Turing's question has thus become a frontier challenge of our time: whether machines can navigate the deep and often irreducible uncertainties of strategic choice.

Central to this challenge is \emph{strategic foresight}: the capacity to form accurate, forward-looking judgments about uncertain, high-stakes business outcomes before they unfold.  Unlike chess or protein folding, where objective functions are fixed and solutions can be verified against independent ground truth, strategic decisions unfold in environments characterized by ambiguity, incomplete information, and endogenous change.  Decision makers must anticipate technological shocks, interpret noisy signals about customer adoption, and act on assessments that cannot be validated until long after the moment of choice.  Strategic decision making is thus an archetype of what scholars have called wicked \citep{churchman_wicked_1967}, ill-structured \citep{simon_structure_1973}, or irreducibly uncertain \citep{knight_risk_1921} problems---domains where prediction requires judgment rather than deduction.

Strategic foresight matters precisely because it is so difficult.  Many core strategy theories hold that above-normal profits and competitive advantage derive from a superior ability to predict the future value of resources, the attractiveness of industries, and value-creation opportunities \citep{csaszar_individual_2018}.  Since strategic decisions pay off in the future, and since advantage stems from either luck or foresight, only foresight is controllable by the firm.  Improving foresight thus gives managers practical levers to shape decision quality and, by extension, firm performance.  Empirical evidence supports this logic: firms exhibiting superior foresight display systematically higher productivity and performance \citep[e.g.,][]{bloom_well_2026}.

Yet human foresight is notoriously limited.  Decades of research show that managers are boundedly rational: they are limited in information processing, constrained in computational capacity, and inconsistent in attention \citep{simon_administrative_1947, kahneman_judgment_1982, kahneman_noise_2016}.  Even seasoned experts often perform no better than chance \citep{tetlock_expert_2005}.  Efforts to improve foresight through broader cognitive representations, teaming, and structured updating have shown promise \citep{tetlock_superforecasting_2015, csaszar_individual_2018, csaszar_distributed_2026, peterson_entrepreneurial_2021, kapoor_peering_2023}.  Yet organizational processes often amplify rather than attenuate individual biases, making disciplined judgment difficult to achieve \citep{powell_behavioral_2011}.

These limitations of human cognition suggest a natural question: might artificial intelligence offer a path forward?  AI systems do not share the same constraints as humans---they have access to vastly more information, possess greater computational capacity, and exhibit unwavering consistency.  Recent work has characterized this potential as ``unbounding rationality'' \citep{csaszar_unbounding_2025}: relaxing the cognitive bottlenecks that have long defined the boundaries of human judgment.  Given the importance of strategic foresight and recent advances in AI, a consequential question emerges: \emph{Can AI outperform humans at strategic foresight?}

Answering this question empirically faces a critical obstacle: most real-world outcomes are already known to both researchers and models, making it difficult to distinguish genuine prediction from pattern retrieval.  Recent work has begun addressing this training-data leakage problem by evaluating LLMs on events occurring after their training cutoffs or on synthetic data not present in training corpora \citep{townsend_are_2023, schoenegger_wisdom_2024, doshi_generative_2025}.  Dedicated forecasting benchmarks have also been developed \citep{karger_forecastbench_2025}.  In strategy research, the closest work is \citet{allen_how_2026}, who test AI strategic decision making inside a business simulation---but simulations test performance against known payoff structures, not foresight about genuinely uncertain outcomes.  Prior work has also examined whether LLMs can replicate expert judgment: \citet{csaszar_artificial_2024} found a 52\% correlation between LLM evaluations and venture capitalist assessments of startup pitches.  Yet correlation with expert opinion leaves open whether either party is correct; absent ground truth, high agreement could reflect shared biases rather than predictive accuracy.  What remains missing is a fully prospective, strategically grounded benchmark in which humans and LLMs make ex ante predictions about live business outcomes under genuine uncertainty---with realized outcomes serving as the criterion for accuracy.

\subsection{Our Approach and Contribution}

We designed a fully prospective prediction tournament using live Kickstarter crowdfunding projects.  Kickstarter provides an ideal setting for studying strategic foresight for several reasons.  First and foremost, this setting captures a canonical strategic decision problem: ranking competing investments under uncertainty using only what is knowable before outcomes are realized, when signals are noisy, multidimensional, and expressed in open-ended narratives \citep[see, e.g.,][]{heshmati_learning_2024}.  Like these processes, the Kickstarter setting requires evaluators to compare multiple proposed ventures and decide which are most likely to succeed before outcomes are known.  Second, the temporal structure permits genuinely prospective evaluation: predictions are made and registered before outcomes are determined.  Third, project success is highly uncertain at the time of evaluation: in our sample, funds raised varied by orders of magnitude, and the task proved difficult even for evaluators with relevant experience.  Finally, outcomes are determined by the market---the collective decisions of thousands of backers---providing an objective ground truth against which predictions can be validated.

To operationalize this setting, we sampled 30 U.S.-based technology projects launched after the training cutoffs of all models studied, capturing each while fundraising was in progress and outcomes were unknowable.  A diverse suite of frontier and open-weight large language models (LLMs)---including GPT-5 variants, Claude 4.5, Gemini 2.5, Grok 4, Gemma 3, Llama 3.1, and DeepSeek 3.2---completed a double round-robin tournament of 870 pairwise comparisons (every project compared against every other project in both presentation orders), producing complete rankings of predicted fundraising success.

We benchmarked these AI-generated forecasts against two human groups.  First, 346 employed U.S. adult managers with experience evaluating business alternatives, recruited via Prolific, completed the identical pairwise comparison task using the same information.  Second, we complemented this with three MBA-trained investors possessing specific investment expertise, who completed full rankings under monitored, no-technology conditions.  After project closure, we compared each evaluator's predicted ranking of fundraising success to the realized outcome ranking based on the total funds each venture actually raised.

Human evaluators achieved rank correlations between 0.04 and 0.45.  Several frontier LLMs exceeded 0.60, with the best---Gemini 2.5 Pro---reaching 0.74.  Read differently, this means the leading model correctly ordered approximately 79\% of venture pairs, while the strongest human forecaster managed roughly 60\%.  These differences are statistically significant and of substantial practical magnitude.  We also examine whether aggregation could improve performance for either AIs or humans, and find essentially no improvement: neither wisdom-of-the-crowd ensembles nor human--AI hybrid teams outperformed the best standalone model.  Finally, we investigate what drives performance differences across LLMs by regressing tournament accuracy against established AI capability benchmarks.

These findings represent a meaningful milestone for strategic decision making.  Just as Deep Blue's 1997 victory demonstrated that machines could surpass humans in a domain once thought to require uniquely human intuition, our results show that frontier LLMs substantially outperform managers in predicting strategically relevant outcomes under identical information conditions.  The criteria for such a demonstration are threefold: (i) a domain of acknowledged human expertise, (ii) machine superiority under fair conditions, and (iii) results difficult to dismiss as narrow or artificial.  Our prospective design, real-world outcomes, and consistent results across multiple metrics satisfy these criteria.  Beyond providing a clean test of human versus machine foresight, our approach offers a template for benchmarking AI capabilities in strategy---and benchmarks matter, as they provide clear measures of performance and guide the development of next-generation AI systems.

Our work advances research on strategic foresight \citep{gavetti_evolution_2016, csaszar_individual_2018, ahuja_managerial_2005, peterson_entrepreneurial_2021, kapoor_peering_2023}, behavioral strategy \citep{powell_behavioral_2011, gavetti_behavioral_2012}, and AI in organizations \citep{brynjolfsson_artificial_2017, krakowski_artificial_2022, agrawal_economics_2019, csaszar_artificial_2024} by demonstrating that LLMs achieve markedly higher predictive accuracy than both managers and investors in a live, high-uncertainty venture-selection setting.

Methodologically, we make two contributions.  First, we introduce a fully prospective benchmarking design that enables clean comparisons of human and machine foresight on economically meaningful outcomes.  Second, we demonstrate AI tournaments as a method for accurately measuring LLM performance when facing multiple discrete alternatives: by decomposing a complex ranking task into elementary pairwise comparisons and aggregating via tournament scoring, we obtain stable orderings that would be difficult to elicit through direct ranking prompts.  Substantively, our results reopen classic debates about the limits of human judgment \citep{simon_administrative_1947, kahneman_thinking_2011} and the role AI may play in strategic decision making.  Finally, we consider the competitive implications of widespread AI adoption, suggesting that while foresight may eventually commoditize, advantage will shift to the complements that shape and act on AI predictions---proprietary data, distinctive problem framing, and organizational capabilities that translate forecasts into timely action and learning.

\section{Theoretical Motivation}

The question of whether AI can outperform humans in strategic foresight lies at the intersection of strategy, decision-making, and artificial intelligence.  This section develops the conceptual foundation for that inquiry.  We first define strategic foresight and review evidence on its human antecedents.  We then develop arguments for why large language models may exhibit superior foresight, followed by counterarguments emphasizing deep uncertainty, theory construction, and the limits of data-driven inference.  These competing perspectives clarify why AI superiority in strategy is theoretically indeterminate---and therefore why prospective empirical tests are necessary.

\subsection{Strategic Foresight}

Strategic foresight---the capacity to form accurate ex ante judgments about uncertain, high-stakes strategic outcomes before they unfold \citep{ahuja_managerial_2005, gavetti_evolution_2016}---is foundational to strategy \citep{cockburn_untangling_2000, csaszar_individual_2018}.  Whether predicting a firm's future performance \citep{peterson_entrepreneurial_2021}, projecting industry evolution \citep{kapoor_peering_2023}, or anticipating the actions of negotiation partners \citep{weber_managers_2024}, foresight serves as the cognitive engine for strategic choice.

This importance stems from the conditions under which firms operate.  Investment horizons are long, information is incomplete, and strategic moves are often irreversible.  When technological trajectories are path-dependent, competitive responses unpredictable, and regulatory environments shifting, errors in foresight compound into persistent performance gaps.  Conversely, accurate expectations allow advantageous positioning, transforming foresight from incidental luck into durable competitive advantage.

Conceptually, strategic foresight can be understood through a Brunswikian lens as representational alignment between a decision-maker's internal model and the latent generative structure of the environment \citep{brunswik_conceptual_1952, csaszar_individual_2018}.  Foresight depends on identifying relevant cues from uncertain environments, weighting them appropriately, and integrating them into cognitive representations that mirror the processes producing future outcomes.

Recent research has begun investigating the drivers of superior and inferior foresight.  \citet{heshmati_learning_2024} found that MBA students with broader mental representations---considering more distinct factors---exhibited superior foresight about venture success on Kickstarter.  \citet{peterson_entrepreneurial_2021} found that entrepreneurs' execution experience actually diminished their foresight about time-to-market.  \citet{kapoor_peering_2023} found that updating beliefs based on new information improved foresight, especially when updating followed Bayesian principles.

This literature has advanced our understanding of individual antecedents of foresight but has focused on variation across human decision-makers rather than across fundamental methods of prediction.  Existing research explains why some managers forecast better than others; it remains silent on how human foresight compares to AI foresight---a comparison that recent technological advances now make tractable.

\subsection{Artificial Intelligence and Strategic Foresight}

Modern artificial intelligence is fundamentally a prediction technology.  While early machine learning mapped inputs to outputs in well-defined numerical domains, large language models (LLMs) extend this capability to text sequences.  By training on vast corpora of human language, these models learn to predict the next word with high accuracy.  Surprisingly, in compressing massive textual data into neural network parameters, capabilities resembling reasoning, abstraction, and understanding appear to emerge \citep{wei_emergent_2022}.  More recent iterations combine word prediction with reinforcement learning from human feedback, and this---together with in-context learning from prompts---has produced extraordinary results: from passing professional examinations to generating novel scientific hypotheses \citep{zhao_survey_2025}.

LLMs thus introduce a forecasting system operating under computational principles fundamentally different from human cognition.  Strategic foresight has historically been studied as a property of boundedly rational individuals and organizations.  LLMs represent a qualitatively new forecasting actor: scalable, internally consistent, and capable of integrating information across orders of magnitude more data than any individual or organization.

From this perspective, the emergence of LLMs reframes a long-standing question in strategy from one about variation among human forecasters to one about competition between forecasting regimes.  This shift raises a natural and consequential research question: can AI outperform humans at strategic foresight?

Existing evidence on AI and strategic decision-making primarily examines LLMs as decision support tools embedded within human processes, not as independent forecasting agents.  For example, experimental work shows that while LLM use can substantially alter the breadth and depth of individuals' mental representations, it does not necessarily improve predictive accuracy in individual strategic decision tasks, particularly under time constraints \citep{kanis_ai-augmented_2026}.  Similarly, \citet{camuffo_beyond_2026} demonstrate that the architecture of the AI system---specifically whether it acts as a routed ``copilot'' versus a general chatbot---determines whether users engage in critical evaluation or passive delegation, with significant downstream effects on confidence and belief updating.  Such findings underscore that AI's impact depends on design, deployment, and human--AI interaction.  By contrast, our study evaluates whether LLMs can independently generate superior forecasts under identical information conditions, abstracting from human--AI interaction effects.

The next two sections develop competing arguments for why AI may, and may not, exceed human strategic foresight.

\subsection{Why AI May Exhibit Superior Strategic Foresight}

Human foresight faces stringent constraints.  Human judgment operates under severe attentional, memory, and computational limits \citep[e.g.,][]{simon_administrative_1947, gigerenzer_bounded_2002}.  Individuals rely on heuristics, analogies, and simplified causal representations when reasoning about complex futures \citep{march_organizations_1955, tversky_judgment_1974}.  These shortcuts generate systematic distortions: base-rate neglect, over-extrapolation from salient cases, and excessive confidence in fragile causal models \citep{kahneman_timid_1993}.

Organizational processes can further degrade foresight.  Hierarchical filtering, political incentives, and local search dynamics, all bias which signals reach decision-makers \citep{cyert_behavioral_1963, levinthal_adaptation_1997, gavetti_origin_2007}.  Even formal forecasting processes suffer from correlated beliefs, shared blind spots, and endogenous feedback between expectations and action \citep{denrell_adaptation_2001, powell_behavioral_2011}.  The limits of human foresight are structural, not merely motivational.

Against this backdrop, AI may relax several foundational constraints that limit human-based strategic foresight.  LLMs have been trained on vast numbers of cases and have abstracted patterns in how humans respond to those cases; they can therefore be used to predict human responses in new situations.  From a Brunswikian perspective, this large-scale exposure may allow LLMs to attend to cues and assign weights that more accurately approximate the latent generative structure of environments than any human decision-maker could achieve \citep{brunswik_conceptual_1952, csaszar_individual_2018}.  This underlies the approach of treating LLMs as \emph{homo silicus}: implicit computational models of human behavior that can be used to simulate and explore social science experiments \citep{horton_large_2023}.  Three mechanisms suggest potential AI superiority.

First, computational capacity.  Classic theories emphasize that humans are boundedly rational: they attend selectively, simplify aggressively, and rely on heuristics \citep{gavetti_looking_2000, denrell_search_2019}.  LLMs can evaluate vast combinatorial spaces, integrate thousands of features simultaneously, and update predictions without fatigue \citep{csaszar_unbounding_2025}.  This advantage matters most where foresight depends on synthesizing many weak, interacting signals---exactly the conditions characterizing strategy.

Second, information scale.  LLMs are trained on massive, heterogeneous corpora spanning technical, commercial, scientific, and cultural domains.  While any human relies on limited experience, LLMs implicitly draw on millions of analogs and market descriptions.  This scale advantage translates into forecasting superiority across domains from elections \citep{argyle_out_2023} to stock prices \citep{lopez-lira_can_2025} to neuroscience \citep{luo_large_2024}.  Aggregating multiple LLMs further improves performance: \citet{schoenegger_wisdom_2024} show that a ``wisdom of the silicon crowd'' rivals elite human forecasting collectives.  Dynamic benchmarks reinforce this pattern \citep{karger_forecastbench_2025, chen_vcbench_2025}.

Third, internal consistency.  Human judgment is not merely biased but noisy: the same individual often gives different answers to identical questions \citep{kahneman_noise_2021, satopaa_decomposing_2021}.  Such noise substantially degrades predictive accuracy.  LLMs produce highly consistent outputs given fixed instructions and low temperature setting, making them well-suited for comparative evaluation tasks where the objective is relative ordering rather than absolute prediction.  Eliminating noise alone can generate large performance gains---a central lesson from the longstanding literature comparing statistical and clinical judgment \citep{meehl_clinical_1954}.\footnote{A related (but more contingent) possibility is bias reduction.  In some workflows, LLM-based support may attenuate biases that often affect human judgment (e.g., anchoring or overconfidence), but LLMs can also inherit and reproduce distortions embedded in training data and prevailing narratives \citep{birhane_algorithmic_2021, bender_dangers_2021}.  Thus, the relevant comparison is not ``biased humans'' versus ``unbiased AI,'' but a shift from idiosyncratic human error toward systematic model error.  Whether that shift improves accuracy should depend on the setting and the magnitude of idiosyncratic error \citep{agrawal_economics_2019, kahneman_noise_2021}.}

Recent strategy research supports these possibilities.  \citet{sen_can_2026} show that LLMs outperform humans in matching structurally correct analogies under competing alternatives.  \citet{allen_how_2026} show that LLMs outperform human decision-makers in stylized strategic simulations with well-defined payoffs.  These studies indicate LLMs can exceed human performance in structured strategic reasoning.  What remains unknown is whether these advantages extend to strategic foresight under genuine uncertainty---where outcomes are unknowable, feedback delayed, and no objectively correct solution exists at the time of choice.

\subsection{Why AI May Not Exhibit Superior Strategic Foresight}

A substantial body of strategy scholarship remains deeply skeptical that AI systems can meaningfully deliver superior strategic foresight.  These critiques do not deny AI's ability to detect patterns in large datasets.  Rather, they argue that strategic foresight is fundamentally not a pattern-recognition problem.  Instead, it is a theory-driven, uncertainty-laden process of constructing novel interpretations, commitments, and courses of action in environments where the relevant data often do not yet exist \citep[e.g.,][]{felin_what_2018, felin_theory_2024}.

A central critique in this vein is that predicting the future requires constructing novel interpretations, understanding causality, and imagining counterfactual futures---capabilities that go beyond mere extrapolation.  As \citet[\p 86]{felin_what_2018} argue, ``Innovative strategies depend more on novel, well-reasoned theories than on well-crunched numbers.''  From this view, AI systems, trained to optimize predictive accuracy over existing corpora, are structurally constrained to operate within the space of already-articulated representations.  They can recombine patterns, but they cannot originate the causal theories that redefine what patterns even matter.  From this perspective, ceding strategic judgment to AI risks reinforcing existing mental models rather than enabling the conceptual reframing that underlies genuine strategic innovation.  If future outcomes depend on such theory formation and reframing, critics argue that AI-based forecasts may systematically miss the inflection points most consequential for strategic foresight.

A second rebuttal to AI's ascendance in strategy emphasizes the deep, Knightian uncertainty that defines many strategic environments.  In such settings, the relevant probability distributions are often unknown or unknowable \citep{knight_risk_1921, simon_structure_1973, gavetti_origin_2007}.  Strategic problems are not merely noisy versions of well-defined statistical tasks; they are often ill-structured, with unclear objectives, latent causal mechanisms, and outcomes endogenously shaped by the strategic actions themselves.  Strategic analogies, frames, and opportunity structures are arguably uniquely constructed through human cognition, not discovered in data \citep{gavetti_looking_2000, gavetti_cognition_2005}.  Because these representations precede and shape what gets measured, critics argue that AI systems (no matter how statistically powerful) remain downstream of the most consequential elements of strategic judgment.  Under such conditions, predictive accuracy depends less on estimating known distributions and more on constructing representations of uncertainty itself, a process that critics argue AI systems may be structurally ill-equipped to perform.

Finally, a third objection stresses that much of what underlies effective strategy (and strategic foresight) is tacit and socially embedded, rather than explicitly codifiable \citep{nelson_evolutionary_1982}.  From this standpoint, strategic mastery depends on experiential patterning, contextual sensitivity, and skilled interpretation that arise through practice rather than computation \citep{levitt_organizational_1988}.  AI systems, lacking embodiment, social immersion, and lived organizational experience, may therefore be unable to access the forms of knowledge that matter most for executive judgment.  If accurate strategic foresight relies on tacit interpretation of weak signals and situated understanding of unfolding contexts, AI systems may systematically misweight or misinterpret precisely the cues that foreshadow future outcomes.

\bigskip

The preceding arguments articulate two competing, internally coherent theories of AI-based strategic foresight.  On one view, AI may outperform humans by relaxing fundamental cognitive bottlenecks, integrating vastly more information, and eliminating inconsistency.  On the other, strategic foresight may remain irreducibly human because it depends on theory formation, tacit knowledge, and judgment under Knightian uncertainty---all capacities that AI systems may be structurally unable to replicate.

These arguments point in opposite directions.  Several mechanisms suggest AI superiority; others suggest decisive human advantage.  Whether AI can outperform humans at strategic foresight is therefore a genuinely open question rather than a foregone conclusion.

Resolving this tension requires direct empirical comparison in an environment capturing the defining features of strategic judgment: forward-looking uncertainty, delayed feedback, and the absence of objectively correct solutions at the time of choice.  Our study provides such a test.

\section{Methods}

Studying strategic foresight with LLMs poses a distinctive challenge: most real-world outcomes are already known to models through their training data.  This training-data leakage problem means that seemingly forward-looking evaluations may actually be retrospective, with models reproducing memorized patterns.  Valid assessment requires evaluating models out-of-sample: on events that had not yet occurred at training time \citep{townsend_are_2023, doshi_generative_2025, schoenegger_ai-augmented_2025, csaszar_unbounding_2025}.

\subsection{Empirical Setting and Sample}

Kickstarter, a prominent online crowdfunding platform where entrepreneurs solicit financial contributions to bring proposed products to market, provides a natural setting for such evaluation.  Each Kickstarter project launches publicly and remains open for several weeks before its fundraising outcome is determined.  By selecting projects initiated after the latest model training cutoffs and capturing their content while funding is still in progress, we ensure that (i) no model could have seen these data during training, and (ii) the outcome variable itself (the ultimate amount raised) has not yet been determined.  This design cleanly avoids any training data leakage, providing a genuinely forward-looking test of predictive capability.

To anonymize and standardize the information set, we provided both LLMs and human forecasters with an approximately 500-word summary of each campaign describing the product, team, key features, and risks, while excluding the venture's name, the platform name (Kickstarter), funds raised to date, consumer comments, and other dynamic signals that might reveal campaign performance.  Testing confirmed that identifying projects via search engines using these summaries is difficult.  This design also guards against cheating: in addition to making search difficult, outcomes could not be looked up because they had not yet been determined.

We sampled the 30 live projects from Kickstarter's Technology category with the closest project-close dates that (i) were U.S.-based, (ii) had funding goals of \$10,000 or more, and (iii) launched after the most recent training cutoff for models we examine (see Appendix~\ref{app:sample} for sample details).  Full-page PDF printouts of each campaign were captured, representing the ex ante information available to potential backers.

\subsection{Prediction Task and Rankings}

Our empirical design evaluates how various LLMs and humans rank these projects based on their ex ante prediction of fundraising success compared to the actual funds raised ex post.  All evaluations and rankings were completed during a 72-hour window (October 31 -- November 2, 2025) while project fundraising was still live and outcomes were not yet known.   We also uploaded the predictions by the LLMs, Prolific participants, and experts to Zenodo in advance of actual outcomes becoming known.\footnote{The preregistration of this study can be found here: \url{https://zenodo.org/records/17501586}}

\subsubsection{Pre-Processing}

To produce uniform inputs, we first applied optical character recognition (using ABBYY FineReader 15) to all PDFs to make text embedded within images accessible, then provided the resulting documents to GPT-5-High, which generated structured summaries of each project (see Appendix~\ref{app:process} for the full prompt and Appendix~\ref{app:summary_output} for an example).  Each summary is approximately 500 words and follows a standard format covering areas such as problem, solution, key features, risks, and team.  Summaries are anonymized: they omit the product name and any reference to Kickstarter.  Pilot testing confirmed that searching Google using reasonable keywords from these summaries rarely surfaced the original project within the first several pages of results.  Summaries excluded all dynamic metrics (reward amounts, backers, time remaining, comments).  These standardized overviews enable comparison across models and human raters on identical information, leveling the playing field for models with limited image recognition.

\subsubsection{LLM Benchmark}

Following common practice in AI benchmarking \citep[e.g.,][]{allen_how_2026, schoenegger_wisdom_2024}, we evaluate a diverse set of frontier and open models differing in architecture, training data, and release date.  We selected a variety of models available when we ran the experiment.  Proprietary models include variants of GPT (5, 5-high, 5-low, 5-mini, 5-nano), Claude (opus-4.1, sonnet-4.5), Gemini (pro-2.5, flash-2.5), and Grok (4, 4-fast-reasoning, 4-fast-non-reasoning).  Open-weight models include Gemma (3-4b, 3-12b, 3-27b), Llama (3.1-8b, 3.1-70b), and DeepSeek (3.2-reasoner, 3.2-chat).  We set temperature to 0.5 for all models permitting this parameter; for the GPT-5 family, which does not expose temperature, we used the model's fixed default.

To generate rankings from pairwise judgments, we employ a double round-robin tournament.  Each of the 30 projects is compared against every other---once in each presentation order (A vs.\ B and B vs.\ A)---yielding 870 total comparisons per model.  This ensures complete coverage of all 435 unique matchups while controlling for order effects, and it replaces an error-prone global ranking with many head-to-head decisions, producing a more stable aggregate ordering from a large set of elementary judgments \citep{david_method_1988}.

For each ordered pair, the model is provided with the two project summaries and asked to determine which would raise the most funds from potential customers and donors.  It is specifically prompted to provide (i) an analysis of the two projects followed by (ii) an explicit statement of which project it expects to win (a forced choice of A or B).  Requesting an analysis before selection follows the chain-of-thought prompting technique, which has been shown to improve accuracy \citep{wei_chain_2022}.  Across all pairings, each project accumulates a win total.  For instance, if Project A is selected as superior in 50 of its 58 ($=29 \times 2$) head-to-head matchups, it receives a higher tournament rank than Project B, which won only 20 of its matchups.

\subsubsection{Human Benchmark: Prolific}

To benchmark LLM performance against human performance in the most comparable manner, we implemented the same tournament described above with human participants.  We recruited 368 fully employed U.S. adults from Prolific with English fluency, managerial experience, and self-reported experience evaluating business alternatives, consistent with prior research \citep{mount_quantum_2021, mickeler_knowledge_2023}.  Participants evaluated the same anonymized summaries in an identical double round-robin design, completing three pairwise comparisons per session (see Appendix~\ref{app:prolific_tournament} for survey text).

Pairings were pre-assigned so no project appeared twice within a participant's six viewed projects, and no participant saw both orders of the same matchup.  This achieved full coverage of 870 comparisons with some redundancy for quality control.

Compensation was \$7.50 plus performance-contingent bonuses: after outcome resolution, five randomly selected participants from the top 10\% of forecasters received bonuses (1st~=~\$100, 2nd~=~\$50, 3rd--5th~=~\$25).  This structure rewards accuracy without encouraging extreme guesses \citep{witkowski_incentive-compatible_2023, schoenegger_ai-augmented_2025}.

Quality controls included attention and comprehension checks.  Materials were anonymized; copy-paste and keystroke events were logged.  Invisible sentinel text in the project descriptions instructing LLMs to use special words, flagged AI-generated content.  Respondents were classified as having used AI assistance if they copy-pasted their answer or typed faster than 100 words per minute.  Respondents flagged for AI use, for copy\&pasting, for typing faster than 100 words per minute, or for failing attention criteria were excluded, yielding 346 participants covering the 870 comparisons.

\subsubsection{Human Benchmark: Experts}

To complement the general manager sample with domain-specific expertise, we recruited three current students or recent graduates of a highly ranked U.S. MBA program with substantial prior professional experience in investment banking, management consulting, venture capital, private equity, and/or corporate development.

Each expert evaluated the same 30 projects as the LLMs and the Prolific participants, using the same project summaries.  To preclude the use of digital assistance, each expert received physical packets containing the project summaries, with each project printed on a separate page to facilitate sorting and annotation.  Experts were explicitly instructed that the task must be completed without the use of AI tools, search engines, or other external technology.  Each expert also joined an individually scheduled, recorded Zoom session while completing the task on camera.

Unlike the double round-robin tournament used for LLMs and Prolific participants, each expert produced a strict complete ordering of all 30 projects from 1 (expected to raise the most funds) to 30 (expected to raise the least), with no ties permitted.  Experts received \$400 for participation, with an additional \$400 performance-contingent bonus awarded to the top-performing expert.

\subsection{Performance Measures}\label{sec:performance-measures}

Strategic foresight concerns the accuracy of predictions about uncertain outcomes.  The economic significance of this capability lies in placing eventual successes above failures, thereby allocating attention or capital efficiently.  Ordering the 30 projects similarly to the realized ranking therefore indicates a high level of strategic foresight.  This is, in fact, very close to Turing's call for an AI CEO who could make ``buy or sell'' decisions.

Our primary dependent variable is Spearman's rank correlation ($\rho$) between each evaluator's predicted ranking and the true ranking based on actual funds raised.  Because $\rho$ is scale-insensitive and penalizes large ranking mistakes more than small ones, it directly operationalizes strategic foresight: getting the order right under uncertainty.

We supplement this with pairwise accuracy---the fraction of project pairs correctly ordered---computed using Goodman-Kruskal's $\gamma$ (excluding tied pairs) and rescaled to $[0,1]$.  This measure is perhaps more tangible than correlation: it represents the probability of picking the correct alternative when facing a pair, and is equivalent to the dependent variable used in related studies \citep{csaszar_individual_2018, heshmati_learning_2024}.  We also measure value capture: the share of total funds raised by the evaluator's top-$K$ predicted projects relative to the true top-$K$.  For value capture, we use fractional allocation when ties occur at the top-$K$ boundary.\footnote{When ties occur at the boundary, each tied project contributes proportionally to the numerator.  For example, if two projects tie for 5th place, each contributes half its raised amount to the top-5 sum.}

\subsection{Uncertainty Quantification}\label{sec:bootstrap}

Because our measures are subject to sampling variability, we require methods to assess whether a given correlation differs significantly from zero or from another evaluator's correlation, and similarly for other dependent variables.  With samples of moderate size such as ours, the recommended approach is nonparametric resampling \citep{good_permutation_2004}.

For each evaluator, we apply project-level nonparametric bootstrap, resampling projects with replacement and recomputing performance measures.  All intervals are two-sided 90\% using $10,000$ replications.  All analyses use a fixed random seed for reproducibility.

\section{Results}

We begin by comparing evaluators' predictive accuracy using Spearman's rank correlation between predicted and realized project rankings.  For brevity, we refer to this simply as ``correlation'' throughout.  We then examine pairwise accuracy and economic value capture to assess the practical significance of these differences.

\subsection{Correlation Results}

\begin{figure}\centering
\includegraphics[width=\linewidth]{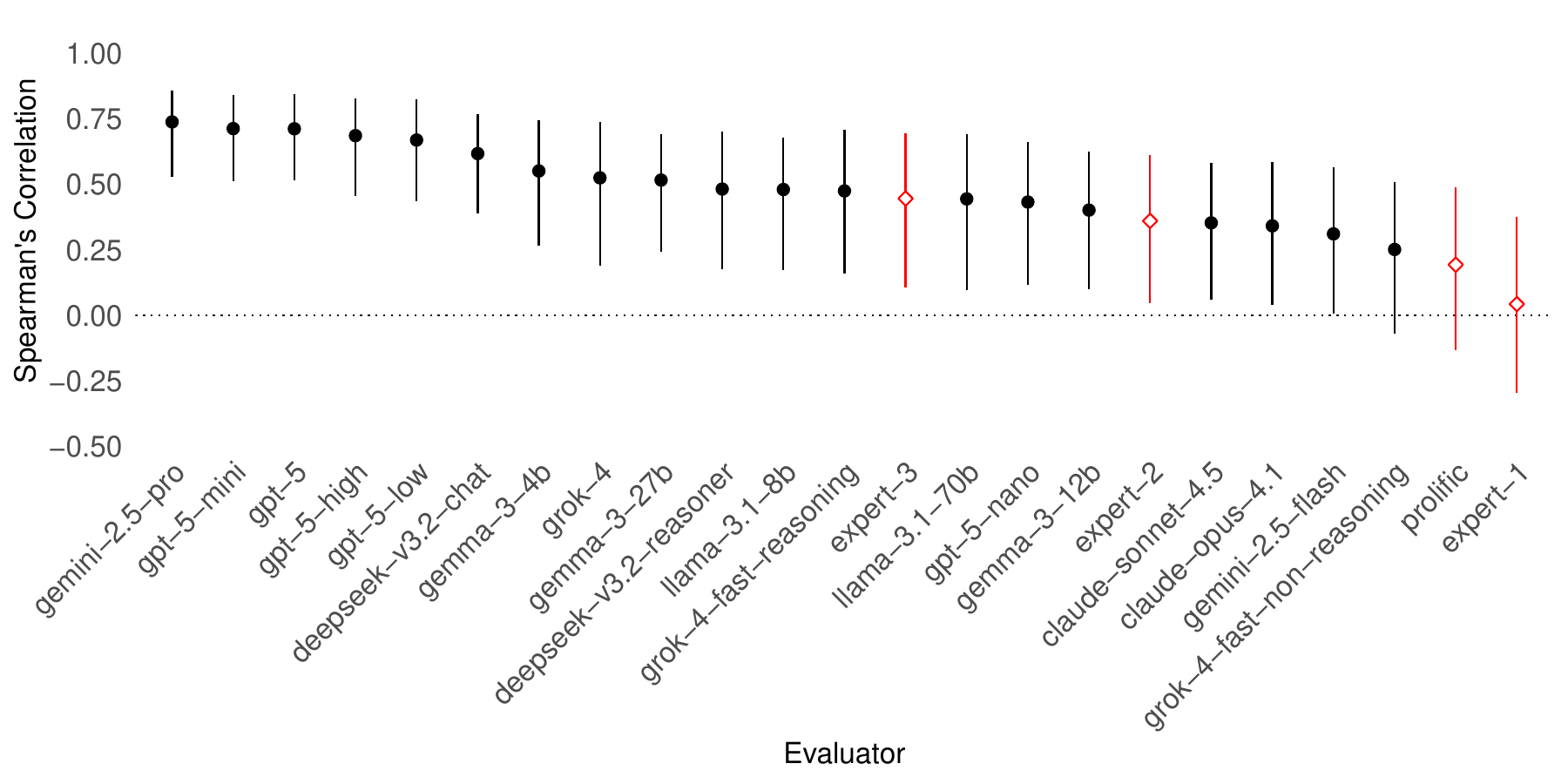}
\caption{Spearman's rank correlation ($\rho$) between predicted and realized project rankings, with 90\% confidence intervals.  The dashed line at zero represents chance performance; human evaluators are shown in blue with diamond markers.}\label{fig:spearman}
\end{figure}

Figure~\ref{fig:spearman} presents the correlation between predicted and realized rankings, with 90\% bootstrap confidence intervals.  A correlation of zero (shown by the dashed line) represents chance---equivalent to random ordering.  Given measurement uncertainty, the minimum correlation required for statistical significance ($\alpha = 0.05$, one-tailed) is approximately 0.31; correlations below this threshold may reflect sampling variability rather than genuine predictive ability.

The results reveal large differences across evaluators.  Gemini 2.5 Pro achieved the highest correlation ($\rho = 0.74$), followed by the GPT-5 family clustering between 0.67 and 0.71.  The best human, Expert~3, achieved 0.45---below many LLMs.  The Prolific crowd ranking produced 0.19, while Expert~1 achieved 0.04; both are statistically indistinguishable from zero.

Human performance exhibited substantial heterogeneity.  Among three MBA-trained investors completing the task under identical conditions, correlations ranged from 0.45 (Expert~3) to 0.04 (Expert~1).  Given the wide confidence intervals, this variation underscores the difficulty of identifying reliable expert judgment in strategic domains where ground truth is unavailable at decision time.

Bootstrap tests on pairwise differences confirm that the top models outperform human benchmarks.  The difference between Gemini 2.5 Pro ($\rho = 0.74$) and Expert~3 ($\rho = 0.45$) is 0.29 ($p = 0.063$).  The difference between Gemini 2.5 Pro and the Prolific crowd ranking ($\rho = 0.19$) is 0.55 ($p = 0.001$).  Full pairwise tests appear in Appendix~\ref{app:significance}.

\subsection{Pairwise Accuracy}

\begin{figure}\centering
\includegraphics[width=\linewidth]{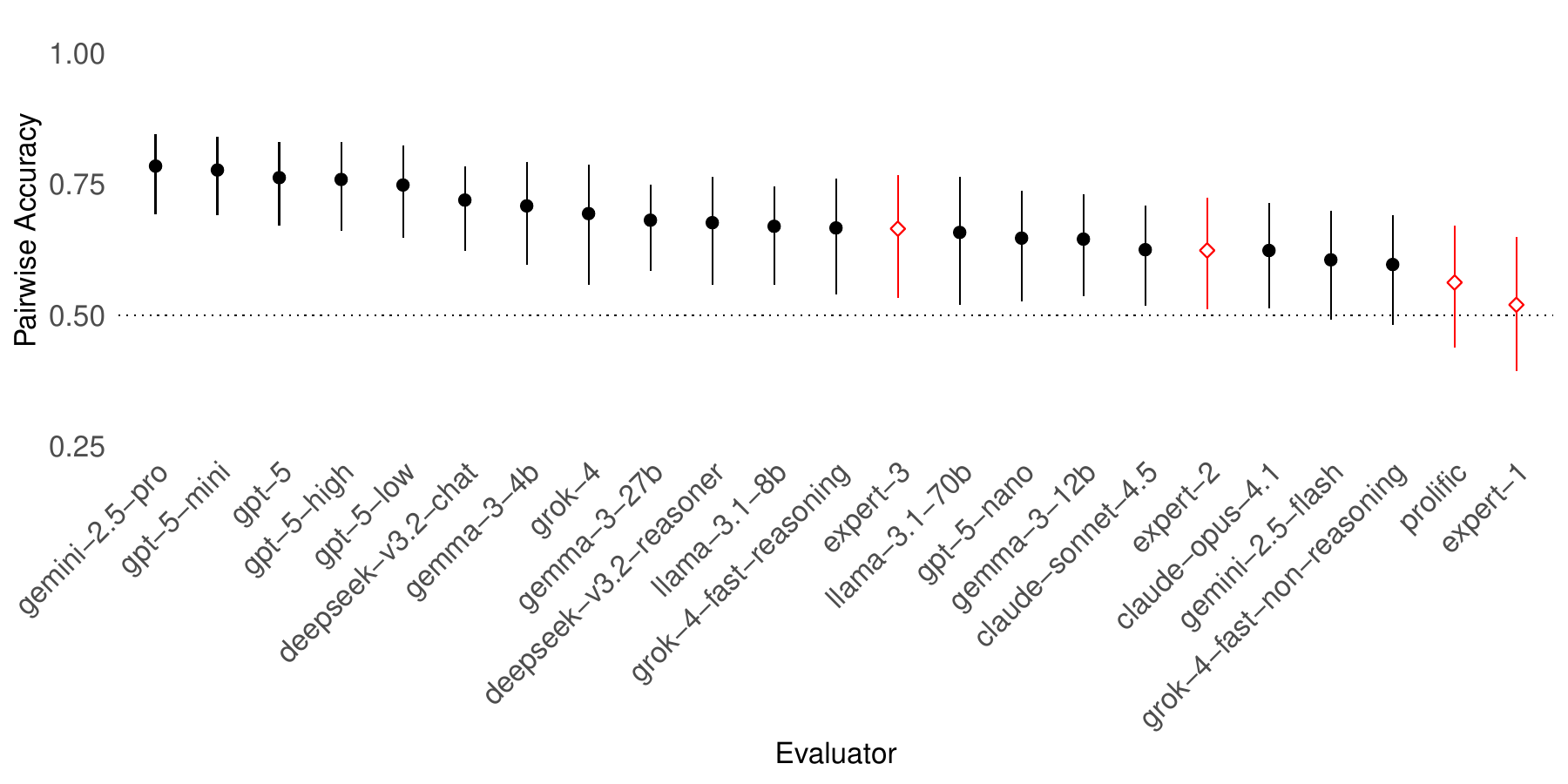}
\caption{Pairwise comparison accuracy, with 90\% confidence intervals.  The dashed line at 50\% represents chance performance; human evaluators are shown in blue with diamond markers.}\label{fig:accuracy}
\end{figure}

Correlation summarizes overall ordinal alignment, but it can be difficult to interpret directly.  To provide a more tangible view of the same performance differences, Figure~\ref{fig:accuracy} presents pairwise comparison accuracy---the fraction of project pairs correctly ordered.  Random guessing yields 50\%.

Consistent with the correlation results, Gemini 2.5 Pro and GPT-5 Mini achieved approximately 79\% and 78\% accuracy---correctly ordering nearly four of five pairs.  Expert~3 achieved 67\%.  The Prolific crowd achieved 56\%, and Expert~1 approximately 52\%; neither is significantly distinguishable from chance.

\subsection{Visualizing Predictive Accuracy}

\begin{figure}\centering
\includegraphics[width=\linewidth]{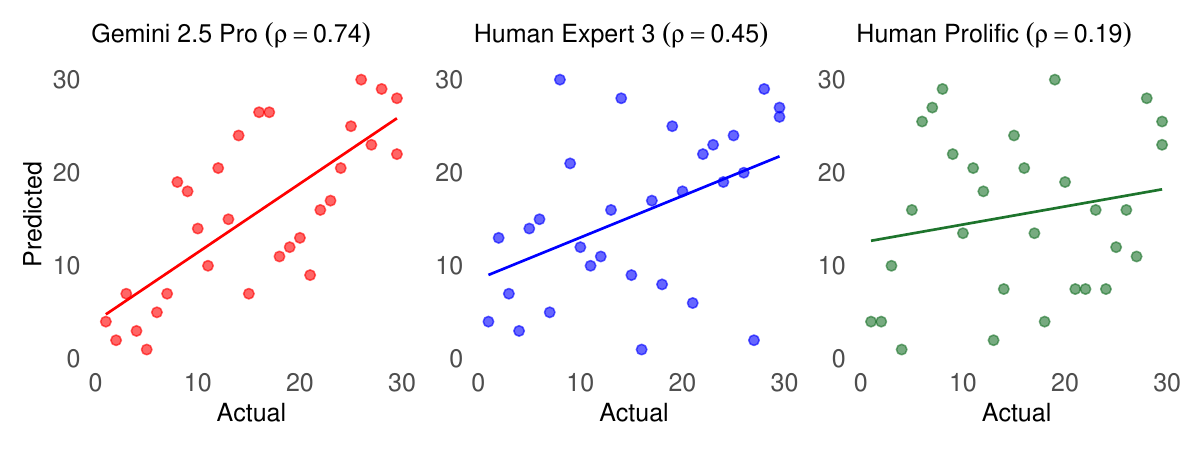}
\caption{Scatter plots of predicted versus actual project rankings for the best LLM (Gemini 2.5 Pro), the best human (Expert~3), and the Prolific crowd ranking.  Lines show linear fit.}\label{fig:scatter}
\end{figure}

Returning to correlations, Figure~\ref{fig:scatter} provides visual intuition for what different correlation magnitudes look like in practice.  The panels show scatter plots of predicted versus actual rankings for the best LLM (Gemini 2.5 Pro), the best human (Expert~3), and the Prolific crowd ranking.

For Gemini 2.5 Pro ($\rho = 0.74$), predictions cluster tightly around the diagonal: projects predicted to succeed generally did.  For Expert~3 ($\rho = 0.45$), the relationship remains positive but with considerable scatter; several highly-ranked projects performed poorly.  For the Prolific crowd ranking ($\rho = 0.19$), scatter appears nearly random.

\subsection{Patterns Across Evaluators}

\begin{figure}[t]
\centering \includegraphics[width=\linewidth]{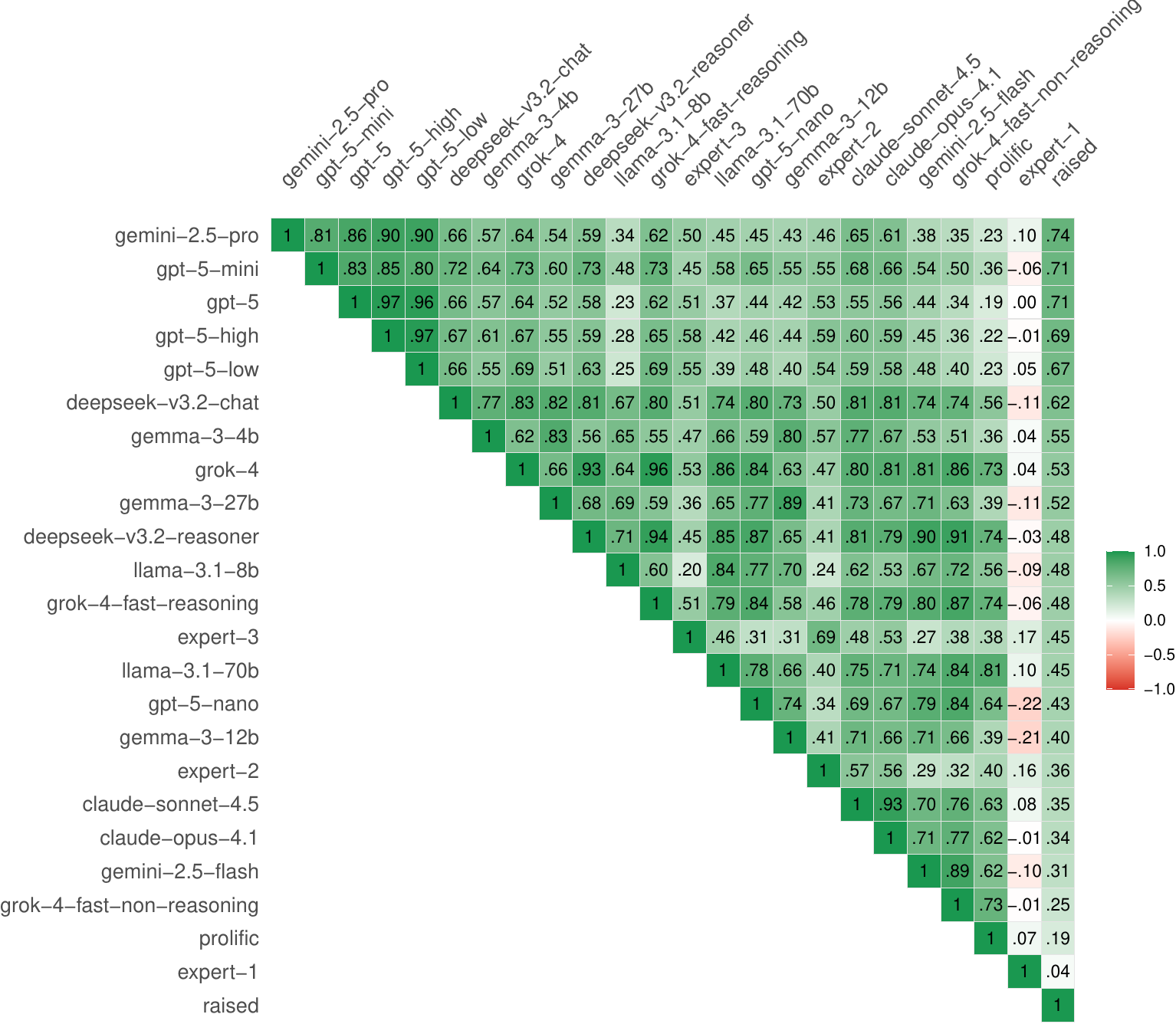}
\caption{Spearman correlation matrix among all evaluators.  The rightmost column shows each evaluator's correlation with actual funds raised.}\label{fig:correlations}
\end{figure}

Figure~\ref{fig:correlations} presents the full correlation matrix among evaluators and with realized outcomes.  The rightmost column, correlation with actual fundraising outcomes, mirrors the ranking hierarchy from Figure~\ref{fig:spearman}.  The rest of the matrix reveals how evaluator rankings correlate with one another.

While our 30-project sample provides power to detect large differences---between frontier LLMs and humans, and between best and worst LLMs (Appendix~\ref{app:significance})---it does not reliably distinguish similarly-ranked evaluators.  However, the overall pattern of differences may illuminate what drives performance variation.

Interpreting these patterns benefits from a brief orientation to the most salient model distinctions in our set.  LLMs differ along several dimensions: model family (reflecting different training data, architectures, and optimization objectives), model size (with smaller models typically representing compressed ``distillations'' of larger siblings), and reasoning capability (with some models explicitly trained to engage in deliberative, step-by-step reasoning).

Models within families correlate highly: GPT-5 variants exhibit inter-correlations of 0.80 to 0.97.  GPT-5-Nano, however, correlates only 0.44 to 0.65 with siblings and performed substantially worse.  A similar pattern appears in Gemini: Pro and Flash share family lineage but differ dramatically in both rankings and outcome correlation.  These within-family gaps suggest that model size matters.

However, model size alone does not predict performance.  Smaller variants sometimes matched or exceeded larger counterparts: Gemma-3-4b outperformed Gemma-3-27b; Llama 3.1-8b slightly exceeded Llama 3.1-70b.  Similarly, explicit reasoning showed inconsistent effects: Grok-4's reasoning variant far outperformed its non-reasoning counterpart, yet DeepSeek Chat outperformed DeepSeek Reasoner.  Claude models ranked near the bottom among LLMs despite being frontier systems according to most common benchmarks (see Table~\ref{tbl:llm_information} in Appendix~\ref{app:sample}).

Human evaluators show low agreement both with one another and with the LLMs.  Correlations between experts and the top models range from roughly 0.2 to 0.58, and inter-expert correlations are similarly modest.  In this setting, then, human rankings appear comparatively unstable across individuals, whereas the top-performing models produce rankings that are not only more accurate with respect to outcomes but also more mutually aligned within families.

\subsection{Economic Significance: Value Capture}

Rank correlations quantify predictive accuracy, but strategic foresight matters insofar as it determines value captured from decisions.  In most strategic contexts, outcomes are highly asymmetric: a few choices yield extraordinary success while many yield limited returns.  This asymmetry is also present in our Kickstarter setting: across the 30 projects, funds raised ranged from \$10.00 to \$481,926.00 (mean \$44,176.23; sd \$115,566.36; see Table \ref{tbl:kickstarter_projects} for details).  In such contexts, the central question is whether an evaluator identifies the exceptional opportunities.

\begin{figure}\centering
\includegraphics[width=\linewidth]{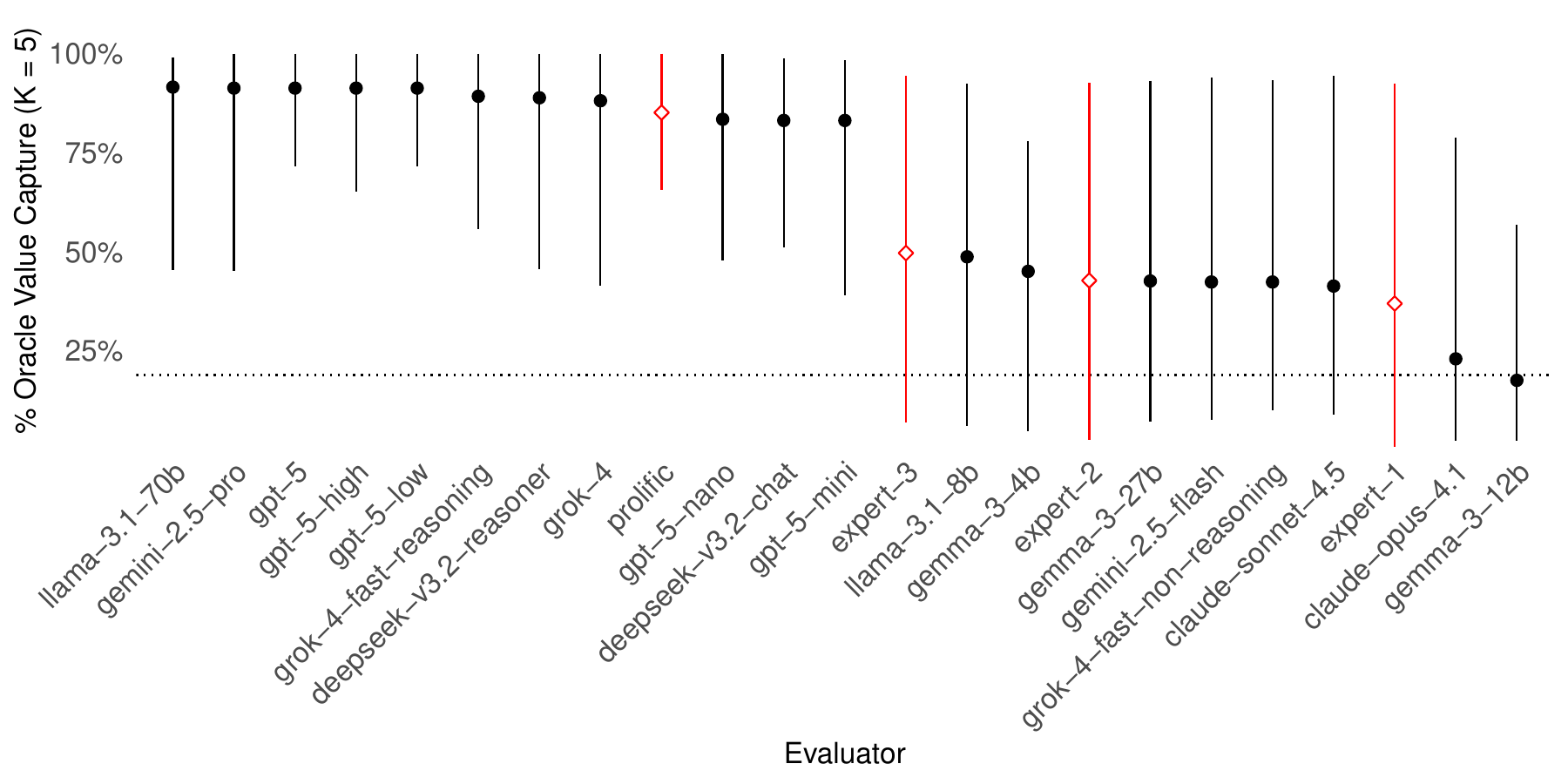}
\caption{Percentage of top-5 project value captured by each evaluator's top-5 predictions, with 90\% confidence intervals.  The dashed line at 18.9\% represents expected capture under random selection; human evaluators are shown in blue with diamond markers.}\label{fig:capture}
\end{figure}

We compute each evaluator's ``value capture''---funds raised by the evaluator's top-$K$ predictions divided by funds raised by the true top-$K$ (the ``oracle'' value).  Figure~\ref{fig:capture} shows value capture for $K = 5$.  The dashed line at 18.9\% represents expected capture under random selection, reflecting outcome skewness.\footnote{Random expected value capture is $K \times \bar{y} / \sum_{i=1}^{K} y_{(i)}$, where $\bar{y}$ is mean outcome and $y_{(i)}$ denotes the $i$th largest outcome.}

Results show a bimodal pattern.  Twelve evaluators---all LLMs or Prolific---captured more than 83\% of oracle value and differ significantly from chance.  All remaining evaluators fell below 50\% and are indistinguishable from random.  Notably, Prolific captured 85\%, outperforming all three experts and representing the only human-based evaluator with statistically significant value capture.

There may also be a gap between reasoning and non-reasoning variants for value capture.  Grok-4 fast-reasoning captured 89\% versus just 42\% for Grok-4 fast-non-reasoning.  Similarly, DeepSeek v3.2 Reasoner (89\%) outperformed DeepSeek v3.2 Chat (83\%).  This suggests that deliberative reasoning may be valuable for identifying outlier opportunities.

\begin{figure}\centering
\includegraphics[width=0.9\linewidth]{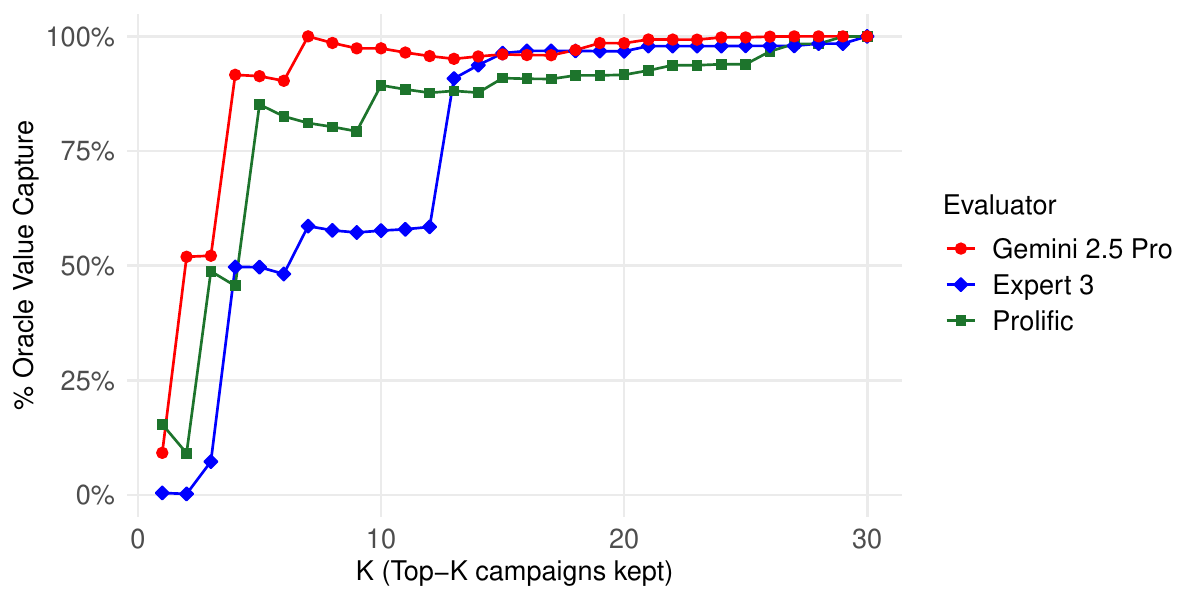}
\caption{Cumulative percentage of total funds captured as portfolio size ($K$) increases from 1 to 30.}\label{fig:topk_curve}
\end{figure}

Figure~\ref{fig:topk_curve} extends this analysis across all values of $K$, plotting cumulative value capture as portfolio size expands.

In this figure, Gemini 2.5 Pro tracks the oracle most closely.  More revealing is the comparison between Prolific and Expert~3. Despite Prolific's lower correlation (0.19 vs.\ 0.45), Prolific captured value more efficiently for portfolios up to $K=12$---reaching 85\% at $K=5$ compared to Expert~3's 50\%.

This apparent paradox---worse overall ranking but better value capture---reflects where errors concentrate.  Rank correlation weights all pairwise comparisons equally: misordering two poor performers counts the same as misordering a poor performer and an exceptional one.  Value capture depends almost entirely on correctly identifying top outliers.  Prolific's errors appear concentrated among comparisons between mediocre projects.  Expert~3, while more accurate on average, made costlier mistakes by failing to rank the highest-value projects at the very top.

This pattern carries a practical implication: evaluators with moderate correlations can still make sound capital allocation decisions if their accuracy is highest where it matters most, at the top of the distribution, where a small number of opportunities drive the majority of returns.  For strategic foresight, not all errors are equally consequential; overall accuracy metrics may matter less than accuracy at the decisive margin.

This pattern does not invalidate correlation as a measure of predictive ability, but it reveals that correlation and value capture answer different questions.  For selecting among frontier LLMs, the two metrics largely agree: Gemini 2.5 Pro, the top performer on correlation, also achieves top-level value capture.  The divergence emerges primarily among weaker evaluators, where moderate overall accuracy can coexist with either good or poor tail performance.  Which metric matters more depends on the decision context: correlation for holistic evaluation quality, value capture for allocation under skewed returns.

\subsection{What Capabilities Drive LLM Performance?}

To shed light on what LLM characteristics drive model performance, we examine how each model's correlation with realized outcomes relates to three established benchmarks spanning distinct capability dimensions of LLMs: MMLU-Pro (broad factual knowledge), Humanity's Last Exam (HLE; expert-level reasoning requiring integration of specialized knowledge with multi-step logic), and LiveCodeBench (procedural and algorithmic competence).  Scores are standardized prior to analysis.  This exercise is exploratory: we aim not to establish causal relationships but to identify which existing capability measures track strategic foresight, thereby informing conjectures about what LLM characteristics may matter most.

Before presenting results, we briefly characterize each benchmark.  MMLU-Pro measures broad academic knowledge and robust reasoning capabilities, analogous to standardized graduate admissions tests (e.g., GRE, MCAT); it assesses whether a model knows what already exists.  Humanity's Last Exam (HLE) tests deep expertise and complex, multi-step logical synthesis, analogous to PhD qualifying examinations; it assesses whether a model can function as a domain expert integrating knowledge across fields.  LiveCodeBench evaluates real-time coding ability and procedural problem-solving, analogous to a technical interview; it assesses algorithmic and implementation competence.

Table~\ref{tbl:benchmark_regression} reports regressions of correlation on standardized benchmark scores.  HLE is the only benchmark with a statistically significant positive association: a one-standard-deviation increase in HLE corresponds to roughly 0.15 increase in correlation.

\begin{table}\centering
\adjustbox{max width=\linewidth}{\begin{tabular}{l d{2.5} d{2.5} d{2.5} d{2.5}}
\toprule
 & \multicolumn{4}{c}{\textit{Dependent variable: Spearman's Correlation ($\rho$)}} \\
\cmidrule(lr){2-5}
 & \multicolumn{1}{p{1in}}{\centering MMLU-Pro}
 & \multicolumn{1}{p{1in}}{\centering HLE}
 & \multicolumn{1}{p{1in}}{\centering LiveCodeBench}
 & \multicolumn{1}{p{1in}}{\centering All Combined} \\
 & \multicolumn{1}{c}{(1)} & \multicolumn{1}{c}{(2)} & \multicolumn{1}{c}{(3)} & \multicolumn{1}{c}{(4)} \\
\midrule
MMLU-Pro (std)         & 0.029          &                &                & -0.037          \\
                       & (0.035)        &                &                & (0.062)         \\
HLE (std)              &                & 0.084^{**}     &                & 0.154^{**}      \\
                       &                & (0.029)        &                & (0.052)         \\
LiveCodeBench (std)    &                &                & 0.045          & -0.051          \\
                       &                &                & (0.034)        & (0.078)         \\
Constant               & 0.512^{***}    & 0.512^{***}    & 0.512^{***}    & 0.512^{***}     \\
                       & (0.034)        & (0.029)        & (0.033)        & (0.028)         \\
\midrule
Observations           & \multicolumn{1}{c}{19}   & \multicolumn{1}{c}{19}   & \multicolumn{1}{c}{19}   & \multicolumn{1}{c}{19}   \\
R$^{2}$                  & 0.040          & 0.323          & 0.092          & 0.439           \\
Adjusted R$^{2}$         & -0.017         & 0.284          & 0.038          & 0.327           \\
\bottomrule
\multicolumn{5}{p{\textwidth}}{\scriptsize\textit{Note:} This table reports regression results where the dependent variable is Spearman's rank correlation between each model's predictions and the outcome (funds raised).  Independent variables are standardized (z-score) benchmark scores.  Standard errors are reported in parentheses.  ***, **, and * indicate significance at the 1\%, 5\%, and 10\% levels, respectively.}
\end{tabular}
}
\caption{OLS regression of Spearman's correlation with realized outcomes ($\rho$) on standardized LLM benchmark scores.  Standard errors in parentheses.}\label{tbl:benchmark_regression}
\end{table}

This pattern is instructive.  HLE tests expert-level reasoning on problems requiring deep, PhD-level knowledge and multi-step logical inference---problems without clear algorithmic solutions.  MMLU-Pro tests broad factual knowledge; LiveCodeBench tests programming ability.  The fact that foresight correlates most closely with HLE suggests that strategic prediction rewards substantive reasoning and cross-domain synthesis over pattern matching or factual retrieval.  Being a ``deep thinker,'' in the sense captured by HLE, appears to drive LLM foresight.  More broadly, this pattern suggests that strategic foresight is a test of synthesis rather than retrieval: knowing facts (MMLU-Pro) or following algorithmic rules (LiveCodeBench) does not predict foresight, but the capacity to connect disparate concepts and reason across domains---the ``PhD-level'' capability captured by HLE---does.

\subsection{Effect of Aggregation}

Given the large differences in individual performance, a natural next question is whether there are simple, low-cost ways to improve accuracy by combining evaluators---either across LLMs, across humans, or through human--AI ``centaur'' combinations.  This is the classic ``wisdom of the crowd'' hypothesis applied to forecasting: aggregation can outperform individuals when errors are not perfectly aligned.  Appendix~\ref{app:aggregation} reports results for several standard aggregation rules (Copeland scores, Borda counts, win summation, and $z$-score aggregation).  The qualitative conclusion is stable across methods: aggregation rarely improves upon the best individual evaluator, and it often reduces performance when weaker evaluators are included.

We find no evidence of a centaur effect under ex post aggregation.  Combining the best human (Expert~3) with the best LLM (Gemini 2.5 Pro) yields $\rho = 0.67$, below the LLM alone.  At the pairwise level, aggregation generally behaves like a compromise: mixing a strong evaluator with a weaker one pulls the combined ranking away from the stronger ordering, lowering correlation with realized outcomes.  The main exception is a combination of two already-strong models: aggregating Gemini 2.5 Pro ($\rho = 0.74$) with GPT-5 Mini ($\rho = 0.71$) yields $\rho = 0.77$, a modest increase of 0.03 that is not statistically significant.

Larger ensembles show the same pattern.  A ``Frontier Trio'' ensemble of the top three models, an ensemble of all 20 LLMs, proprietary-only and open-weight-only ensembles, a human-only ensemble, and a grand ensemble combining all evaluators all fall below the best individual model (Gemini 2.5 Pro).  The implication is straightforward: for this task and information set, selecting the single best model dominates simple averaging and voting schemes.  These findings do not rule out gains from interactive human--AI collaboration during forecasting, but they indicate that naive aggregation of independent predictions is not a free performance upgrade.

\subsection{Robustness}

Appendix~\ref{app:robustness} reports a preregistered replication on a second, independent sample of Kickstarter projects constructed using the same selection procedure as the initial sample.  This robustness check evaluates LLM performance only (not Prolific participants or the three experts).  After excluding three projects removed from the platform, 26 projects remained.  The results reveal qualitatively similar patterns across models.

The robustness sample constitutes a more challenging prediction task due to greater outcome compression: 30.8\% of projects raised less than \$100 (compared to 20.0\% in the initial sample), and 15.1\% of project pairs differed in realized funds by \$100 or less (compared to 5.3\% initially).  This compression makes accurate ordinal prediction intrinsically more difficult.  Consistent with this increased difficulty, absolute correlations are slightly lower.  However, the relative ordering of model performance remains moderately stable: across the 19 models appearing in both samples, the correlation between performance rankings is approximately 0.54.  Models that performed best initially---including GPT-5 variants and Gemini 2.5 Pro---remain among the strongest performers despite lower absolute correlations, providing additional support for our main conclusions.

\section{Discussion}

Can AI outperform humans at strategic foresight?  In a fully prospective, preregistered study, frontier LLMs substantially outperformed both experienced managers and MBA-trained investors given identical information and aligned accuracy objectives.  Several LLMs achieved rank correlations nearly twice as large as the best human evaluators and captured a disproportionate share of economic upside.  Put differently, when the task is to choose, repeatedly, between pairs of competing ventures, the best-performing LLM selected the ultimately higher-performing venture roughly four times in five; the strongest human managed roughly three in five.  Even modest-seeming differences of this magnitude imply large downstream effects when screening decisions are repeated across many investments.

\subsection{A Chess Moment for Strategy?}

These findings invite comparison to the ``chess moment'' of 1997, when Deep Blue's victory over Kasparov demonstrated that machines could exceed human performance in a domain long considered to require intuition, pattern recognition, and strategic depth.  Our results share the essential structure of that demonstration: a fully prospective test, identical information conditions, and a clear performance gap favoring AI.  In both cases, the result challenges a prior assumption---that certain forms of judgment remain uniquely human.

Yet the analogy has limits.  Chess is a single, definite game; by contrast, strategic foresight spans heterogeneous decisions across industries, time horizons, and competitive contexts \citep{simon_structure_1973}.  We examined one such context---a ``fruit fly'' setting amenable to controlled study---not the full population of strategic problems.  Moreover, while Kasparov represented an unambiguous pinnacle of human chess performance, no equally settled reference point exists for strategic foresight; it is unclear who the world's best strategic evaluator is or how one would identify them.  We therefore do not claim that the chess moment for strategy has arrived---only that it no longer appears distant or impossible.  In at least one economically meaningful prediction task under genuine uncertainty, frontier AI systems substantially outperformed experienced human evaluators.  Whether this generalizes across the broader landscape of strategic decisions remains open.  The frontier has shifted; how far it will continue to move is a question this study poses but does not answer.

\subsection{Adjudicating Between Competing Theories}

Our theoretical development articulated two competing perspectives on AI-based strategic foresight.  One view emphasized mechanisms by which AI could relax the structural sources of human forecasting error---limitations in information integration, computational capacity, and internal consistency.  The opposing view argued that strategic foresight is constrained by Knightian uncertainty, the construction of novel interpretations, and tacit, situated judgment \citep[e.g.,][]{felin_what_2018, felin_theory_2024}.  Our evidence supports the first perspective for the particular strategic function our task isolates: evaluation.

The observed pattern aligns with the logic of AI advantage.  Frontier models produce markedly higher correlations with the actual outcomes than managers and investors, while human performance is heterogeneous and, for some evaluators, indistinguishable from chance.  Moreover, the association between tournament performance and a reasoning benchmark (Humanity's Last Exam) rather than a broad knowledge benchmark (MMLU-Pro) is consistent with foresight depending on structured inference rather than retrieval alone.  Notably, this level of performance is difficult to reconcile with characterizations of LLMs as mere ``stochastic parrots'' or next-token predictors \citep{bender_dangers_2021}.  If LLMs were simply reproducing statistical regularities from training data without any capacity for structured inference, we would not expect them to substantially outperform experienced human evaluators on genuinely novel ventures---projects that did not exist during training and whose outcomes were determined by future market responses.  The observed performance gap suggests that whatever computational processes underlie LLM predictions, they capture something more than surface-level pattern matching.

The skeptical view, however, retains a plausible interpretation.  Critics have argued that strategic foresight depends on theory construction, tacit knowledge, and judgment under deep Knightian uncertainty \citep{felin_what_2018, felin_theory_2024}.  Our results do not refute this claim; rather, they may indicate that our task environment does not strongly implicate these mechanisms.  Kickstarter ventures present standardized information, relatively short feedback cycles, and outcomes determined by aggregated consumer response---conditions that may be more amenable to pattern-based inference than, say, predicting the success of a novel business model in an emerging industry.  The relevant question thus becomes one of boundary conditions: in which task environments does AI-based foresight excel, and in which does it falter?  Delimiting these boundaries constitutes a central agenda for future research.

\subsection{Practical Implications}

Our findings apply most directly to contexts structurally similar to our setting: early-stage screening decisions where evaluators rank opportunities based on standardized information under genuine uncertainty, with outcomes determined primarily by external market response rather than by the evaluator's own implementation capabilities.  Venture scouts reviewing deal flow, corporate development teams triaging acquisitions, and innovation committees prioritizing R\&D projects all face analogous tasks.  In such contexts, frontier LLMs may offer immediate value as first-pass filters, particularly where volume exceeds human attention capacity.\footnote{For early evidence along similar lines, see \citet{csaszar_artificial_2024} and \citet{qu_role_2026}.}

The performance differences translate into concrete operational gains.  LLMs correctly ordered approximately 79\% of project pairs versus 60\% for the strongest human.  Where returns concentrate in outliers---as in venture investing and our setting, where funds raised ranged from \$10 to nearly \$500,000---accuracy at the top of the distribution matters most.  Several LLMs captured over 83\% of top-5 value; the best human captured 50\%.  Consistency may prove equally valuable: LLMs produced stable rankings while human evaluators exhibited high variance, with expert correlations ranging from 0.04 to 0.45 under identical conditions.

Our finding that naive human--AI aggregation does not improve upon the best standalone model should be interpreted as a warning about an \emph{augmentation trap}.  When machine performance already exceeds human baselines by a wide margin on a well-defined evaluation task with standardized information, inserting human judgment as a final filter can degrade performance by reintroducing noise and idiosyncratic error.  This pattern may not hold in settings with richer information environments, longer time horizons, or outcomes that depend substantially on the evaluator's own actions (e.g., post-acquisition integration).  Yet even where the augmentation trap applies, it does not imply that humans have no role.  In settings like ours, the most valuable human contributions may lie before and after evaluation: framing the right questions, assembling the relevant information, and translating forecasts into action.

Beyond accuracy, this shift to AI-based foresight offers an advantage in \emph{auditability}.  Human strategic intuition is often opaque and retrospective---a ``gut feel'' that is difficult to challenge.  Because the models in our tournament produced reasoning traces (via chain-of-thought) prior to selection, the ``logic'' of the prediction becomes an organizational artifact that can be inspected, debated, and refined.  This transforms foresight from a private cognitive act into a public organizational resource.

If LLM-based foresight diffuses widely, the strategic question shifts from ``who has better predictions?'' to ``who has better inputs to predictions?''  Our finding that LLMs outperform humans given identical information implies that advantage will increasingly derive from proprietary data, distinctive problem framing, and the organizational capabilities to act on forecasts faster than competitors.  The locus of advantage may migrate from the prediction itself to its complements.

\subsection{Limitations and Future Research}

Several limitations bound our claims.  Kickstarter fundraising is a meaningful but incomplete proxy for venture performance; some strategic dimensions---long-run technological shifts, organizational design, multi-actor competition---are not captured.  LLMs were evaluated on standardized summaries rather than the full, noisy information environments managers face.  Our sample of thirty projects, while sufficient to detect large performance differences between frontier LLMs and human evaluators, limits statistical power to distinguish among similarly performing models.  These limitations invite rather than constrain further research: the conditions under which AI-based foresight exceeds, matches, or falls short of human judgment remain largely unmapped.

A first set of extensions concerns \emph{methods} for AI-augmented foresight.  Our design holds the information set largely fixed by providing standardized summaries and uniform prompts.  Future work should examine how variation in information structure and prompt design affects foresight quality.  This question of \emph{representation}---how AI systems should filter and combine incoming information---may become a central design task in AI-augmented strategy.  A distinct but equally critical methodological challenge concerns \emph{aggregation}.  Our results show that naive aggregation yields limited gains.  This invites more disciplined analysis: under what conditions does combining multiple AI predictions add value, and when does it merely average correlated errors?  Under what conditions can humans and AI together make better predictions than either alone?  The relevant question is not whether aggregation is beneficial in general, but when it restores independence of errors rather than diluting the strongest signal.

A second set of extensions concerns \emph{settings}.  Demonstrating AI superiority in one domain does not establish generality; the strength of any such claim depends on replication across diverse contexts.  Future work should therefore test other industries, tasks, and time horizons.  Three extensions seem particularly important.  First, environments where strategic interaction is central: settings in which one actor's choices induce reactions by competitors, regulators, or other stakeholders, creating feedback loops absent from our single-shot prediction task.  Second, settings where part of the prediction is internal rather than purely external: in contexts such as acquisitions, a material component of the outcome depends on the focal firm's ability to realize synergies, integrate cultures, or execute operational changes---not merely on how the external market will respond.  Our Kickstarter setting isolates external market response; many strategic decisions require predicting both external reactions and internal execution capacity.  Third, settings with longer time horizons and greater novelty, where the patterns in training data may be less informative about future outcomes.  Extending prospective benchmarks to such mixed internal--external outcomes would sharpen understanding of what, precisely, AI can and cannot forecast.

A final set of extensions concerns how to study strategic foresight in environments that lack the convenient properties of our ``fruit fly'' setting, where feedback is rapid and public.  In many strategic domains, outcomes are noisy, delayed, and rare, making direct measurement of predictive accuracy costly or impractical.  One potential avenue is to leverage the logic of robotics development, where agents learn to predict environmental responses in high-fidelity simulations before transferring that knowledge to the physical world.  Just as robotics relies on accurate physics engines to enable ``sim-to-real'' transfer, advancing strategic foresight may require the development of ``market engines''---simulations that preserve the causal structure of competition and allow models to practice prediction without the costs of real-world failure.  Alternatively, research should seek to identify which computational benchmarks reliably track strategic foresight.  Our results suggest that reasoning-heavy benchmarks (such as Humanity's Last Exam) may correlate with foresight better than knowledge-retrieval tasks.  Systematically identifying such proxies would allow researchers to optimize strategic models against stable metrics, reducing reliance on costly and delayed field testing.  More broadly, the emergence of LLMs offers a route to making strategy computable: once strategic judgment can be elicited from AIs, scored against realized outcomes, and revised in light of that evidence, the strategic decision-making process becomes a self-correcting domain rather than a contest of conviction.

\subsection{Conclusion}

Eighty years after Turing dismissed the ambition of building a powerful brain in favor of something ``like the President of the American Telephone and Telegraph Company,'' we arrive at an unexpected juncture.  In a fully prospective prediction tournament---where outcomes were unknown and no model could have seen the data during training---frontier LLMs substantially outperformed experienced managers and MBA-trained investors at precisely the task Turing envisioned: digesting facts about uncertain ventures and providing the judgment to answer ``Do I buy or sell?''

The theoretical contribution here is specific but consequential.  We demonstrate that the cognitive bottlenecks emphasized by bounded rationality theory---limited information integration and inconsistency---are binding constraints on human foresight that AI can relax.  When strategic foresight reduces to evaluation under standardized information, AI superiority is not merely possible but demonstrable.  Whether strategy formulation in uncertain environments constitutes an irreducible source of human advantage, or whether it too will yield to scaled computation, remains the central open question our findings pose.

Methodologically, we introduce \emph{prospective benchmarking}---evaluation on outcomes that are genuinely unknowable at assessment time---as a design template for strategy research.  This approach resolves the training-data leakage problem that plagues retrospective AI evaluations and provides a replicable framework for tracking the evolving frontier of machine capability.

Practically, our results reveal an \emph{augmentation trap}: the intuition that human oversight improves AI decisions may reverse when machine performance already exceeds human baselines.  The value of human judgment lies not in filtering AI outputs but in shaping AI inputs---framing questions, curating data, and designing the organizational architectures that translate predictions into action.  If our results generalize, the strategic landscape will bifurcate: firms that treat AI foresight as a commodity will compete on execution; firms that cultivate it through proprietary data and distinctive framing may find new sources of durable advantage.  The question Turing posed in 1943 has an answer, at least in part.  The question that now confronts strategy scholars is what to build on that answer.

\clearpage
\bibliography{refs}

\clearpage
\appendix
\section{Information on projects and models}\label{app:sample}

Table~\ref{tbl:llm_information} includes the list of 19 LLMs used in the study along with descriptive information.

\begin{table}[!htb]
\centering
\adjustbox{max width=\textwidth}{\begin{tabular}{@{}lllccc@{}}
\toprule
\textbf{Model}            & \textbf{Producer} & \textbf{Parameters}                     & \textbf{MMLU-Pro} & \textbf{HLE} & \textbf{LiveCodeBench} \\
\midrule
gemma-3-27b               & Google            & temp: 0.5                               & 0.669             & 0.047        & 0.137                  \\
gemma-3-12b               & Google            & temp: 0.5                               & 0.595             & 0.048        & 0.137                  \\
gemma-3-4b                & Google            & temp: 0.5                               & 0.417             & 0.052        & 0.112                  \\
llama-3.1-70b             & Meta              & temp: 0.5                               & 0.676             & 0.046        & 0.232                  \\
llama-3.1-8b              & Meta              & temp: 0.5                               & 0.476             & 0.051        & 0.116                  \\
claude-opus-4.1           & Anthropic         & temp: 0.5                               & 0.88              & 0.119        & 0.654                  \\
claude-sonnet-4.5         & Anthropic         & temp: 0.5                               & 0.875             & 0.173        & 0.714                  \\
deepseek-v3.2-chat        & DeepSeek          & temp: 0.5                               & 0.862             & 0.105        & 0.593                  \\
deepseek-v3.2-reasoner    & DeepSeek          & temp: 0.5                               & 0.837             & 0.222        & 0.862                  \\
gemini-2.5-flash          & Google            & temp: 0.5                               & 0.832             & 0.111        & 0.695                  \\
gemini-2.5-pro            & Google            & temp: 0.5                               & 0.862             & 0.211        & 0.801                  \\
gpt-5                     & OpenAI            &                                         & 0.867             & 0.235        & 0.703                  \\
gpt-5-high                & OpenAI            & model: gpt-5 \& reasoning\_effort: high & 0.871             & 0.265        & 0.846                  \\
gpt-5-low                 & OpenAI            & model: gpt-5 \& reasoning\_effort: low  & 0.860              & 0.184        & 0.763                  \\
gpt-5-mini                & OpenAI            &                                         & 0.837             & 0.197        & 0.838                  \\
gpt-5-nano                & OpenAI            &                                         & 0.772             & 0.076        & 0.763                  \\
grok-4                    & xAI               & temp: 0.5                               & 0.866             & 0.239        & 0.819                  \\
grok-4-fast-reasoning     & xAI               & temp: 0.5                               & 0.850              & 0.170         & 0.832                  \\
grok-4-fast-non-reasoning & xAI               & temp: 0.5                               & 0.730              & 0.050         & 0.401                  \\
\bottomrule
\end{tabular}}
\caption{Large language models included in the study, with identifiers, temperature settings, and benchmark scores. All benchmark scores were pulled from Artificial Analysis in December 2025.}\label{tbl:llm_information}
\end{table}

Table~\ref{tbl:kickstarter_projects} includes the list of projects included in the main analysis along with descriptive information.

\begin{sidewaystable}
\centering
\scriptsize
\begin{tabularx}{\textwidth}{@{}llllrrX@{}}
\toprule
\textbf{ID}    & \textbf{Subcategory} & \textbf{Deadline} & \textbf{Launch} & \textbf{Goal} & \textbf{Raised} & \textbf{URL} \\
\midrule
toolkaiser     & 3D Printing          & 11/01/25          & 10/01/25             & \$10,000      & \$12,869        &
\href{https://kickstarter.com/projects/gridmat/toolkaiser}{\nolinkurl{kickstarter.com/projects/gridmat/toolkaiser}} \\
kebo           & Gadgets              & 11/01/25          & 09/17/25             & \$50,000      & \$715           &
\href{https://kickstarter.com/projects/kebottleopener/kebo}{\nolinkurl{kickstarter.com/projects/kebottleopener/kebo}} \\
kujietool      & Fabrication Tools    & 11/02/25          & 09/14/25             & \$10,000      & \$138,547       &
\href{https://kickstarter.com/projects/564580014/kujietool-the-compact-mill-for-makers}{\nolinkurl{kickstarter.com/projects/564580014/kujietool-the-compact-mill-for-makers}} \\
genaix         & Web                  & 11/05/25          & 10/06/25             & \$25,000      & \$125           &
\href{https://kickstarter.com/projects/1327323563/genaix-20-building-the-future-of-human-centered-ai-learning}{\nolinkurl{kickstarter.com/projects/1327323563/genaix-20-building-the-future-of-human-centered-ai-learning}} \\
hply           & Web                  & 11/06/25          & 10/07/25             & \$35,000      & \$6,604         &
\href{https://kickstarter.com/projects/546557384/hply-smarter-giving-made-simple}{\nolinkurl{kickstarter.com/projects/546557384/hply-smarter-giving-made-simple}} \\
pictomic       & Camera Equipment     & 11/09/25          & 09/08/25             & \$49,900      & \$2,998         &
\href{https://kickstarter.com/projects/pictomic/pictomic-macro-photography-rail-slider}{\nolinkurl{kickstarter.com/projects/pictomic/pictomic-macro-photography-rail-slider}} \\
humanecheck    & Apps                 & 11/07/25          & 10/08/25             & \$25,000      & \$13,172        &
\href{https://kickstarter.com/projects/496302772/humanecheck-scan-barcodes-for-animal-welfare-grades}{\nolinkurl{kickstarter.com/projects/496302772/humanecheck-scan-barcodes-for-animal-welfare-grades}} \\
ubo-pod        & DIY Electronics      & 11/07/25          & 10/08/25             & \$25,000      & \$37,200        &
\href{https://kickstarter.com/projects/ubopod/ubo-pod-hackable-personal-ai-assistant}{\nolinkurl{kickstarter.com/projects/ubopod/ubo-pod-hackable-personal-ai-assistant}} \\
dreamie        & Hardware             & 11/07/25          & 10/07/25             & \$10,000      & \$44,358        &
\href{https://kickstarter.com/projects/ambient/dreamie}{\nolinkurl{kickstarter.com/projects/ambient/dreamie}} \\
pickup         & Apps                 & 11/07/25          & 10/08/25             & \$15,500      & \$15,600        &
\href{https://kickstarter.com/projects/pickupapp/pickup-the-basketball-app}{\nolinkurl{kickstarter.com/projects/pickupapp/pickup-the-basketball-app}} \\
angelry        & Wearables            & 11/08/25          & 09/09/25             & \$20,000      & \$21,821        &
\href{https://kickstarter.com/projects/1687452435/angelry-the-worlds-first-cultivated-leather-jewelry}{\nolinkurl{kickstarter.com/projects/1687452435/angelry-the-worlds-first-cultivated-leather-jewelry}} \\
soloist        & Apps                 & 11/08/25          & 10/07/25             & \$10,000      & \$11,165        &
\href{https://kickstarter.com/projects/soloistapp/soloist-infinite-loop-pedal}{\nolinkurl{kickstarter.com/projects/soloistapp/soloist-infinite-loop-pedal}} \\
cop-rights     & Apps                 & 11/08/25          & 09/24/25             & \$15,000      & \$25            &
\href{https://kickstarter.com/projects/jersey/what-cops-can-and-cannot-do}{\nolinkurl{kickstarter.com/projects/jersey/what-cops-can-and-cannot-do}} \\
zulni          & Apps                 & 11/09/25          & 09/25/25             & \$10,000      & \$282           &
\href{https://kickstarter.com/projects/fwa-zulni/zulni-journaling-with-biofeedbacks-for-calm-and-clarity-app}{\nolinkurl{kickstarter.com/projects/fwa-zulni/zulni-journaling-with-biofeedbacks-for-calm-and-clarity-app}} \\
voyagx         & Apps                 & 11/12/25          & 10/13/25             & \$28,000      & \$35            &
\href{https://kickstarter.com/projects/voyagxapp/voyagx-where-travelers-connect}{\nolinkurl{kickstarter.com/projects/voyagxapp/voyagx-where-travelers-connect}} \\
clockchain     & Software             & 11/13/25          & 10/14/25             & \$44,500      & \$50            &
\href{https://kickstarter.com/projects/315841049/clockchain-proving-whats-real-online}{\nolinkurl{kickstarter.com/projects/315841049/clockchain-proving-whats-real-online}} \\
my-story       & Apps                 & 11/13/25          & 10/14/25             & \$250,000     & \$808           &
\href{https://kickstarter.com/projects/specialprincess/my-story-my-way}{\nolinkurl{kickstarter.com/projects/specialprincess/my-story-my-way}} \\
hooper         & Apps                 & 11/14/25          & 10/15/25             & \$10,000      & \$10            &
\href{https://kickstarter.com/projects/1748255141/hooper-app-to-find-games-rank-up-and-connect-hoopers}{\nolinkurl{kickstarter.com/projects/1748255141/hooper-app-to-find-games-rank-up-and-connect-hoopers}} \\
lifesummary    & Apps                 & 11/14/25          & 10/15/25             & \$35,000      & \$8,773         &
\href{https://kickstarter.com/projects/797876113/lifesummaryai-your-lifes-most-important-documents}{\nolinkurl{kickstarter.com/projects/797876113/lifesummaryai-your-lifes-most-important-documents}} \\
fraction       & Hardware             & 11/14/25          & 09/30/25             & \$15,000      & \$428,684       &
\href{https://kickstarter.com/projects/deglace/fraction-the-worlds-first-modular-vacuum-built-to-last}{\nolinkurl{kickstarter.com/projects/deglace/fraction-the-worlds-first-modular-vacuum-built-to-last}} \\
cyberdesk      & Software             & 11/14/25          & 09/15/25             & \$20,000      & \$2,245         &
\href{https://kickstarter.com/projects/cyberdefensedesk/cyberdefensedesk}{\nolinkurl{kickstarter.com/projects/cyberdefensedesk/cyberdefensedesk}} \\
remotion-ai    & Apps                 & 11/14/25          & 10/15/25             & \$10,000      & \$10            &
\href{https://kickstarter.com/projects/remotionai/remotion-ai}{\nolinkurl{kickstarter.com/projects/remotionai/remotion-ai}} \\
lumivisor      & Gadgets              & 11/14/25          & 10/15/25             & \$15,000      & \$20,907        &
\href{https://kickstarter.com/projects/lumivisor/lumivisor-flexible-led-film-for-helmet-visors}{\nolinkurl{kickstarter.com/projects/lumivisor/lumivisor-flexible-led-film-for-helmet-visors}} \\
lockguard      & Hardware             & 11/15/25          & 10/09/25             & \$25,000      & \$752           &
\href{https://kickstarter.com/projects/lockguard/lockguard-smart-security-that-travels-with-you}{\nolinkurl{kickstarter.com/projects/lockguard/lockguard-smart-security-that-travels-with-you}} \\
rent-a-bee     & Apps                 & 11/18/25          & 10/14/25             & \$60,000      & \$573           &
\href{https://kickstarter.com/projects/1252482382/rent-a-bee-bridging-nature-and-technology}{\nolinkurl{kickstarter.com/projects/1252482382/rent-a-bee-bridging-nature-and-technology}} \\
prolo-ring     & Hardware             & 11/18/25          & 10/15/25             & \$10,000      & \$481,926       &
\href{https://kickstarter.com/projects/prolo/prolo-ring-precision-control-for-keyboard-power-users}{\nolinkurl{kickstarter.com/projects/prolo/prolo-ring-precision-control-for-keyboard-power-users}} \\
viaia          & Apps                 & 11/21/25          & 10/07/25             & \$13,200      & \$14            &
\href{https://kickstarter.com/projects/viaia/viaia-connecting-travelers-with-local-experiences}{\nolinkurl{kickstarter.com/projects/viaia/viaia-connecting-travelers-with-local-experiences}} \\
sale-finder    & Apps                 & 11/21/25          & 09/22/25             & \$50,000      & \$195           &
\href{https://kickstarter.com/projects/salearoundthecorner/the-sale-around-the-corner-map-it-find-it-flip-it}{\nolinkurl{kickstarter.com/projects/salearoundthecorner/the-sale-around-the-corner-map-it-find-it-flip-it}} \\
robotin-r2     & Hardware             & 11/22/25          & 09/23/25             & \$10,000      & \$74,373        &
\href{https://kickstarter.com/projects/robotin/robotin-r2-worlds-first-robot-carpet-cleaner}{\nolinkurl{kickstarter.com/projects/robotin/robotin-r2-worlds-first-robot-carpet-cleaner}} \\
bugout-battery & Hardware             & 11/29/25          & 09/30/25             & \$24,997      & \$451           &
\href{https://kickstarter.com/projects/bugoutbattery/not-a-toy-bugoutbattery-all-in-one-off-grid-solar-power}{\nolinkurl{kickstarter.com/projects/bugoutbattery/not-a-toy-bugoutbattery-all-in-one-off-grid-solar-power}} \\
\bottomrule
\end{tabularx}

\caption{Kickstarter projects included in the main sample, with funding goals, amounts raised, campaign dates, and URLs.}
\label{tbl:kickstarter_projects}
\end{sidewaystable}

\section{Prompts}\label{app:process}

\subsection{Prompt to Create Project Summaries}\label{app:summary}

\begin{lstlisting}[title={Prompt:}]
**Task:** Produce a 400-500-word detailed summary of a crowdfunding campaign page shown in the attached file.

**Focus & Constraints**

* Summarize what the creators communicate through text and visuals.  Do **not** describe colors, layout, or graphic design.
* Describe only functions, claims, and specifics explicitly stated; do **not** infer or invent features or technical capabilities.
* Exclude all live campaign metrics (e.g., reward tiers, pledge amounts or goals, number of backers, time remaining, comments, similar projects).
* **Do not include the name of the company, product, or any individuals.** Replace all proper nouns (including partners/suppliers, institutions, brands, and handles) with generics like "the company," "the product," "the founders," or "a partner."  Do not include URLs, social handles, or other identifiers.
* **Do not mention the source platform or any website.** Do not state or imply that the information comes from Kickstarter or any other site.  If necessary, refer generically to "the campaign" or "the project page."
* Use Markdown.  Do **not** include any preamble or commentary---begin immediately with the campaign narrative.

**Structure (use these headers, in order; omit any section that isn't explicitly covered on the page):**

1. **## One-Sentence Summary**
   A concise, plain-language statement of what the product is and the core value proposition.

2. **## Problem**
   The pain point or gap the creators say they are addressing, including any context or evidence they provide.

3. **## Solution / Product Overview**
   What the product is and how it addresses the problem at a high level.

4. **## Key Features & Capabilities**
   Bullet or short paragraphs listing the stated features, functions, and outcomes (only those explicitly claimed).

5. **## Target Users & Use Cases**
   Who it's for and primary scenarios of use as described by the creators.

6. **## Development Status**
   What exists today (prototype, beta, production-ready), testing completed, and any concrete proof points shown or stated.

7. **## Roadmap & Delivery Plan**
   Milestones the team commits to (e.g., testing, certification, manufacturing prep, software releases).  Use only stated stages and timing.

8. **## Why Crowdfunding / Use of Funds**
   The stated purpose for raising funds and how funds will be applied (non-monetary descriptions only; no amounts or tiers).

9. **## Risks & Challenges**
   The main delivery risks the team acknowledges and any mitigations they outline.

10. **## Team & Partnerships**
    Who is building it, relevant experience, and partners or suppliers if mentioned (use generics per the redaction rule).

11. **## What's Included**
    Contents of the product/app/access (e.g., items in the box, software components, access entitlements), as explicitly listed.

12. **## Technical Details**
    Any concrete specifications or requirements the creators state (e.g., dimensions, materials, compatibility, OS versions, interfaces).  Include only what's explicitly provided.

13. **## Support & Post-Launch Plans**
    Promised support, updates, warranties, documentation, community channels, or content update cadence, if stated.

**Style Rules**

* Keep each section concise and evidence-based; omit any header with no explicit statements to support it.
* Preserve the creators' ordering of ideas when possible, but follow the header order above for clarity.
* Use short paragraphs; bullets are fine for features/specs if the campaign uses lists.
* Quote only when necessary for clarity; otherwise paraphrase neutrally **and maintain all redaction rules**.
* Do not include hyperlinks or external references---and redact any unique identifiers.
\end{lstlisting}

\subsection{Sample Project Summary}\label{app:summary_output}

\begin{lstlisting}[title={Output:}]
## One-Sentence Summary
A donation app and website that connects people with vetted nonprofits, makes giving fast and secure, and provides clear updates showing the impact of each contribution.

## Problem
The creators say many people want to help but hesitate because it's hard to know which organizations are trustworthy, where the money goes, and whether it truly helps; the process often feels long and ends in unwanted emails and data sharing, which creates friction and reduces support for worthy causes.

## Solution / Product Overview
The product is a donation platform designed to reduce doubt and complexity. It connects users to a curated set of organizations that have been screened for transparency, accountability, and results, enables quick, secure donations, and delivers follow-up updates so donors can see how their generosity made a difference.

## Key Features & Capabilities
* **Vetted organizations:** Every listed nonprofit is reviewed for transparency, accountability, and proven outcomes.
* **Smart discovery:** Tools help users find causes aligned with their values and interests.
* **Simple, secure giving:** A streamlined flow aims to remove "endless forms" and confusing steps, enabling donations in under a minute.
* **Real-time impact updates:** Donors receive updates from supported organizations to understand the effects of their gifts.
* **Privacy protections:** The team commits to no spam, no data selling, and no pressure, limiting communications to gratitude and impact updates selected by the user.
* **Transparent fees:** The platform states a flat 1% fee, with no hidden charges.
* **Mobile and web access:** The experience is available via a mobile app and a web platform.

## Target Users & Use Cases
The product is pitched to both new donors and people who already give regularly. Primary scenarios include discovering trusted organizations (from large, global efforts to local, community causes), making quick donations with confidence, and staying connected through ongoing impact updates.

## Development Status
According to the page, the platform is built, nonprofits are already vetted, and the mission and product scope are defined; the creators present it as ready to expand to more users.

## Roadmap & Delivery Plan
Planned work includes launching the web and mobile platform at scale, expanding outreach to additional nonprofits and prospective donors, and building tools that deliver more meaningful impact insights to users. Specific dates are not listed.

## Why Crowdfunding / Use of Funds
Funds are intended for launch activities, nonprofit and user outreach, and development of impact-insight tools that enhance donor feedback after giving.

## Risks & Challenges
The team identifies technical and operational challenges related to scaling for a broad user and cause base. They state prior startup and product-deployment experience, say the platform is built for flexibility and scalability, and commit to continued transparency as they grow.

## Team & Partnerships
The project is led by the founders, who cite experience leading people operations at a technology company that later went public, with a mission background in safety and impact. The platform features a mix of large international organizations and local community groups as partners, all described generically and subject to vetting.
\end{lstlisting}

\subsection{Prompt to Compare two Projects}\label{app:llm_tournament}

\begin{lstlisting}[title={Prompt:}]
You are an expert crowdfunding analyst tasked with evaluating two projects to determine which will raise the most funds from potential customers and donors.

## Your Task:
Analyze the two projects below and determine which will raise the most funds from potential customers and donors.

## Evaluation Criteria:
Assess each project based on the following factors:

1. **Problem-Solution Fit**: How compelling is the need being addressed?  Is it a real pain point?
2. **Market Analysis**: What is the potential market size?  Who is the target audience?
3. **Competitive Landscape**: What alternatives exist?  What is this project's unique value proposition?
4. **Feasibility**: How realistic is the delivery timeline?  Does the team have credible expertise?
5. **Funding Appeal**: How attractive is this to backers (emotional appeal, rewards, social impact)?
6. **Risk Assessment**: What are the potential red flags or challenges?

## Response Format:
- Provide a structured analysis of **approximately 300 words**
- Compare both projects across the key criteria
- Support your conclusion with specific evidence from the project descriptions
- Your response MUST end with your final verdict as the absolute last line, using this **exact format**:

**[I BELIEVE THE WINNER WILL BE PROJECT X]**

(where X is either 1 or 2)

---

# Project 1:
...

# Project 2:
...

---

**Begin your analysis now.**
\end{lstlisting}

\section{Prolific Crowdworker Survey}\label{app:prolific_tournament}

\begin{singlespacing}
\subsection*{Part 1: Overview}

Welcome!  Before beginning, we'll ask a few brief questions to confirm eligibility for this study.

\medskip \noindent \textbf{Evaluating Strategic Opportunities}

You are being asked to participate in a research study.  Scientists do research to answer important questions that might help change or improve the way we do things in the future.  This document will give you information about the study to help you decide whether you want to participate.  Please read this form, and ask any questions you have before agreeing to be in the study.

All research is voluntary.  You can choose not to take part in this study.  If you decide to participate, you can change your mind later and leave the study at any time.  You will not be penalized or lose any benefits if you decide not to participate or choose to leave the study later.

This research is intended for individuals 18 years of age or older.  If you are under age 18, do not complete the survey.

The purpose of this study is to examine how individuals evaluate strategic opportunities.  We are asking you if you want to be in this study because you are a U.S. resident, with English as your primary language, are currently have full-time employment, and have at least a bachelors degree.

If you agree to be in the study, you will take part in a one-time online survey on the Qualtrics platform.  After reading this information sheet and clicking the forward arrow to indicate consent, you will evaluate three unique pairs of startup ventures (reviewing 6 of 30 brief summaries) and indicate which venture you believe is more promising in each pair.  You will also provide short written explanations for your reasoning.  You will answer a total of 6 questions.  The amount of time commitment to complete the survey will take approximately 15 minutes and will conclude after this one-time survey.

Before agreeing to participate, please consider the risks and potential benefits of taking part in this study.

You may be uncomfortable while answering the survey questions.  While completing the survey, you can skip any questions that make you uncomfortable or that you do not want to answer.

There is a risk someone outside the study team could get access to your research information from this study.  More information about how we will protect your information to reduce this risk is below.

We don't think you will have any personal benefits from taking part in this study, but we hope to learn things that will help researchers in the future.  In particular, we anticipate that this research will help decision makers within organizations make better strategic decisions in the future.

You will be compensated \$7.50 for participation.  Additionally, following the resolution of the forecasting questions in December 2025, a random selection of 5 of the top 10\% forecasters in terms of accuracy (that assesses how closely your probabilistic forecasts align with the collective wisdom of the crowd) will receive up to \$100 in additional rewards based on their performance relative to other participants (1st =\$100, 2nd place = \$50, 3rd - 5th place = \$25).

We will protect your information and make every effort to keep your personal information confidential, but we cannot guarantee absolute confidentiality.  No information which could identify you will be shared in publications about this study.

Your personal information may be shared outside the research study if required by law.  We also may need to share your research records with other groups for quality assurance or data analysis.  These groups include the Indiana University Institutional Review Board or its designees, and state or federal agencies who may need to access the research records (as allowed by law).

If you have questions about the study or encounter a problem with the research, contact the researcher, Daniel Wilde at wilded@iu.edu.

For questions about your rights as a research participant, to discuss problems, complaints, or concerns about a research study, or to obtain information or to offer input, please contact the IU Human Research Protection Program office at 800-696-2949 or at irb@iu.edu.

\subsection*{Part 2: Demographics and Screening}

\noindent \textbf{In your current role, which responsibilities do you regularly perform?  (Select all that apply)}
\begin{itemize}[noitemsep,topsep=0pt,label=$\Box$]
\item Team scheduling / staffing (e.g., shift planning, hiring allocation)
\item Budget approval of $\ge\$5,000$ (e.g., expense or project budgets)
\item Evaluation of alternative business opportunities (e.g., products, suppliers, ventures)
\item Preparing compliance documentation (e.g., internal policies, regulatory forms)
\item Managing workplace safety programs (e.g., OSHA, incident prevention)
\item None of the above
\end{itemize}

\medskip

\noindent \textbf{Which of the following tools have you used in the past 12 months?  (Select all that apply)}
\begin{itemize}[noitemsep,topsep=0pt,label=$\Box$]
\item Slack
\item Asana
\item Jira
\item Zoom
\item AngelList
\item None of the above
\end{itemize}

\medskip

\noindent \textbf{Which best describes your current role level?  (Select one)}
\begin{itemize}[noitemsep,topsep=0pt,label=$\ocircle$]
\item Individual contributor (no management)
\item Manager (any level)
\item Senior leadership (Director/VP/C-suite/founder)
\item Student / not employed
\item Other
\end{itemize}

\subsection*{Part 3: Evaluation}

\noindent \emph{Important Research Policy:} This study must be completed entirely on your own, without the use of AI tools (e.g., ChatGPT, Claude, Gemini, Bard, Copilot, or similar assistants).  Responses are automatically monitored for patterns associated with AI use.

\noindent Analyze the two projects below and determine which will raise the most funds from potential customers and donors.

\noindent Assess each project based on the following factors:

\begin{enumerate}[noitemsep,topsep=0pt]
\item \emph{Problem-Solution Fit}: How compelling is the need being addressed?  Is it a real pain point?
\item \emph{Market Analysis}: What is the potential market size?  Who is the target audience?
\item \emph{Competitive Landscape}: What alternatives exist?  What is this project's unique value proposition?
\item \emph{Feasibility}: How realistic is the delivery timeline?  Does the team have credible expertise?
\item \emph{Funding Appeal}: How attractive is this to backers (emotional appeal, rewards, social impact)?
\item \emph{Risk Assessment}: What are the potential red flags or challenges?
\end{enumerate}

\medskip \noindent \textbf{Project A}
[Summary of Project A]

\medskip \noindent \textbf{Project B}
[Summary of Project B]

\medskip

\noindent Provide a structured analysis comparing both projects across the key evaluative criteria mentioned above.  Support your conclusion with specific evidence from the project descriptions.  (minimum of 50 words)

\medskip

\noindent Which of the two projects will raise the most funds from potential customers and donors?
\begin{itemize}[noitemsep,topsep=0pt,label=$\ocircle$]
\item Project A (shown first above)
\item Project B (shown second above)
\end{itemize}

\medskip

\noindent [Repeated two more times for a total of 3 comparisons]

\end{singlespacing}

\section{Statistical Significance Tests}\label{app:significance}

Because all evaluators predict the same set of project outcomes, their predictions are not independent.  As such, standard comparisons of non-overlapping confidence intervals can be misleading.  To rigorously test whether performance differences are statistically significant, we bootstrap the pairwise difference in Spearman's $\rho$ between evaluators directly.
For each pair of evaluators $(i, j)$, we resample the 30 projects with replacement 10,000 times.  In each bootstrap iteration, we compute $\rho_i$ and $\rho_j$ on the resampled data and record their difference $\Delta\rho = \rho_i - \rho_j$.  We then construct two-tailed 90\% confidence intervals from the bootstrap distribution and compute $p$-values.

\begin{figure}[!tb]
\centering
\includegraphics[width=\linewidth]{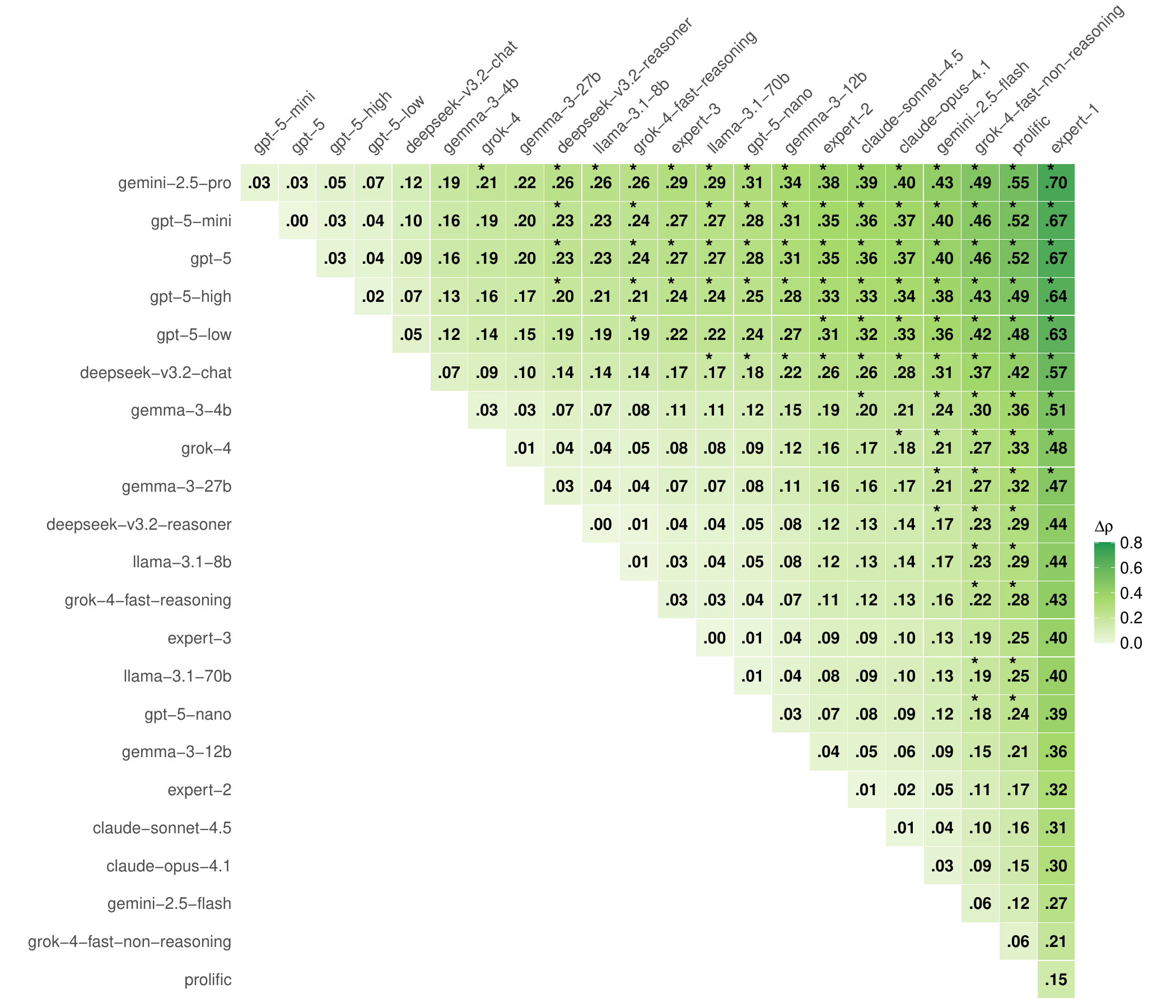}
\caption{Pairwise differences in Spearman's $\rho$ between evaluators (row minus column).  Asterisks indicate statistical significance at the 90\% level.}
\label{fig:significance_heatmap}
\end{figure}

Figure~\ref{fig:significance_heatmap} presents the full matrix of pairwise differences.  Each cell shows $\Delta\rho$ (row evaluator minus column evaluator), with asterisks indicating significance at the 90\% level.

Several patterns emerge from these tests.  First, the top-performing LLMs (e.g., Gemini 2.5 Pro, GPT-5 family) significantly outperform human evaluators at conventional levels.  The difference between Gemini 2.5 Pro and Expert 3 ($\Delta\rho = 0.29$, $p = 0.063$) is significant at the 90\% level despite our modest sample of 30 projects.  Second, differences among top LLMs are not statistically significant: Gemini 2.5 Pro versus GPT-5 ($\Delta\rho = 0.03$, $p = 0.69$) and GPT-5 versus GPT-5 Mini ($\Delta\rho \approx 0$, $p = 0.96$) cannot be distinguished.  Third, while Expert 3 outperforms Prolific in point estimates ($\Delta\rho = 0.25$), this difference is not statistically significant ($p = 0.18$), reflecting the high variance in human judgment.

Overall, of the 253 unique pairwise comparisons, 96 (37.9\%) are statistically significant at the 90\% level.  The significance pattern broadly follows the performance ranking: high-performing LLMs are significantly better than low-performing LLMs and humans, while differences among similarly-ranked evaluators are generally not significant.

\section{Aggregation and Ensemble Methods}\label{app:aggregation}

A natural question is whether combining predictions from multiple evaluators---either multiple LLMs, multiple humans, or human-AI ``centaur'' teams---can improve forecasting accuracy beyond the best individual.  We explore this using several aggregation methods.

\subsection{Aggregation Methods}

We considered four approaches to combining rankings: Borda counts (assigning points based on rank position), Copeland scores (counting pairwise victories across evaluators), win summation (aggregating raw pairwise comparison wins), and $z$-score aggregation (standardizing each evaluator's scores before combining).  We report results using $z$-score aggregation, which normalizes each evaluator's win counts to zero mean and unit variance before summing across evaluators.  This approach accounts for differences in scale and spread across evaluators and enables fair combination of LLM pairwise comparison data with expert forced rankings.  All four methods yield substantively similar results.

\subsection{Pairwise Aggregation Results}

\begin{figure}[!tb]
\centering
\includegraphics[width=\linewidth]{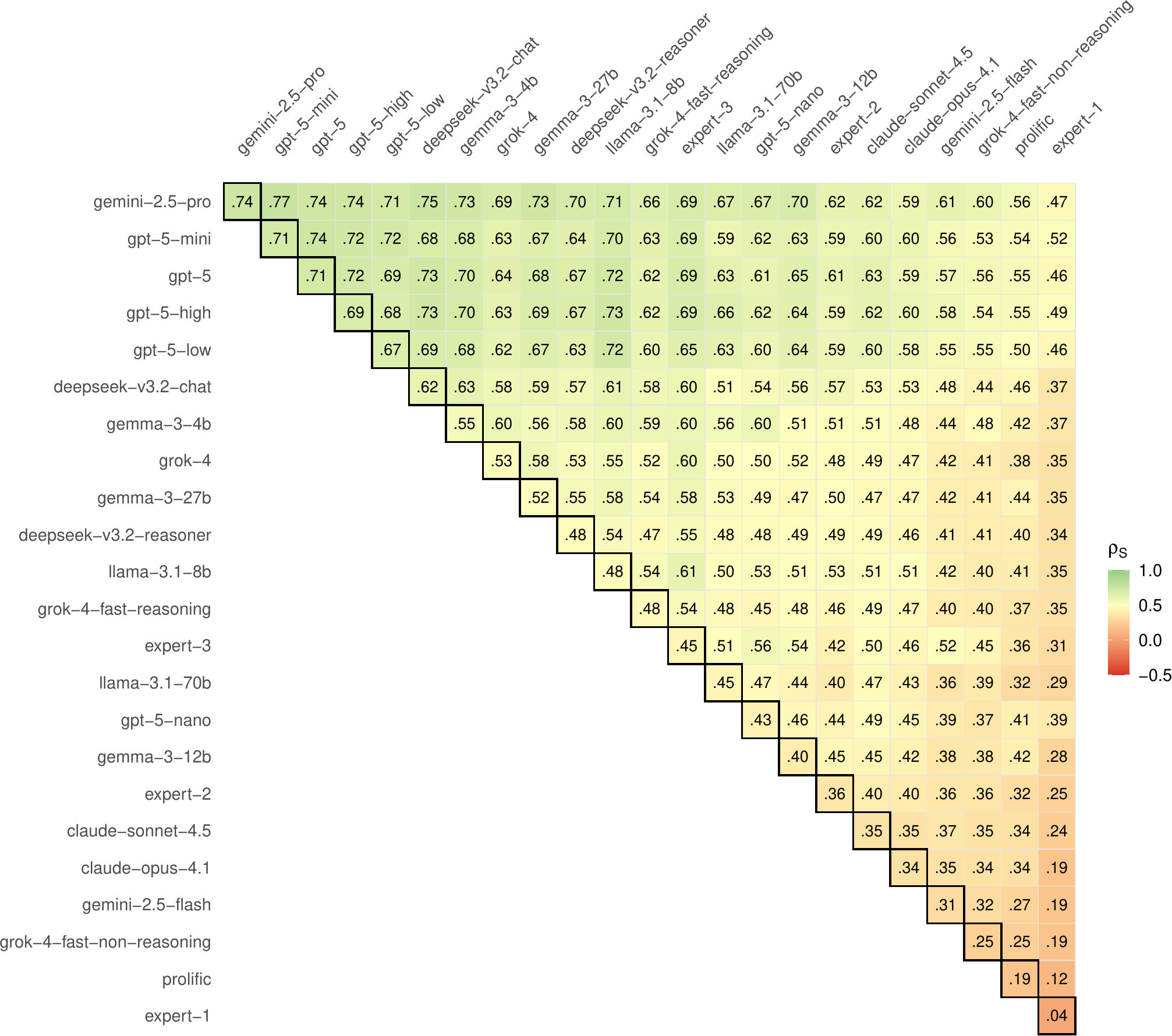}
\caption{Spearman correlation with realized outcomes for pairwise aggregates using $z$-score averaging.  Diagonal cells (outlined) show individual performance; off-diagonal cells show aggregate performance.}
\label{fig:aggregate_heatmap}
\end{figure}

Figure~\ref{fig:aggregate_heatmap} shows the correlation with realized outcomes when aggregating each pair of evaluators using $z$-score averaging, with diagonal entries showing individual performance and off-diagonal entries showing the aggregate.  The key finding is that aggregation rarely improves upon the best individual in each pair.  Only one combination---Gemini 2.5 Pro plus GPT-5 Mini ($\rho = 0.74$ and $\rho = 0.71$, respectively)---achieves a higher correlation ($\rho = 0.77$) than the best individual model alone, a modest and statistically insignificant improvement of 0.03.  In most cases, aggregating a strong evaluator with a weaker one reduces performance relative to using the strong evaluator alone.

\subsection{Ensemble Performance}

\begin{figure}[!tb]
\centering
\includegraphics[width=0.8\linewidth]{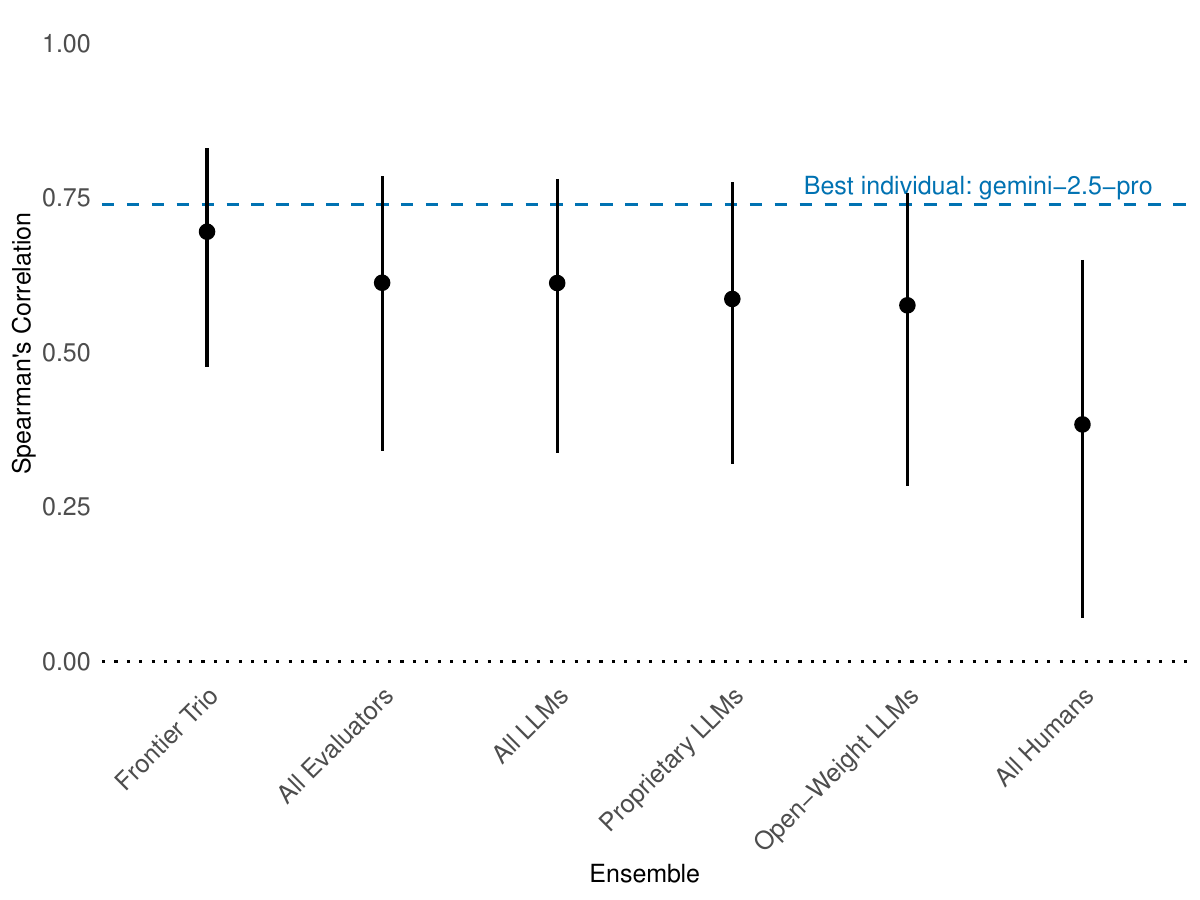}
\caption{Spearman correlation with realized outcomes for ensemble strategies versus the best individual evaluator (Gemini 2.5 Pro, dashed line), with 90\% confidence intervals.}
\label{fig:ensemble}
\end{figure}

Figure~\ref{fig:ensemble} compares various ensemble strategies against the best individual evaluator (Gemini 2.5 Pro).  We construct ensembles for several groupings: a ``Frontier Trio'' of the three best-performing models (Gemini 2.5 Pro, GPT-5 Mini, GPT-5), all 20 LLMs, proprietary LLMs only (GPT, Gemini, Claude, Grok), open-weight LLMs only (Llama, Gemma, DeepSeek), all human evaluators, and all evaluators combined.  No ensemble outperforms the best individual.  The Frontier Trio ensemble achieves $\rho \approx 0.69$, below Gemini 2.5 Pro's $\rho = 0.74$, and ensembles that include weaker evaluators perform progressively worse.

\subsection{No Centaur Effect}

We find no evidence of complementarity between human and AI predictions in ex post aggregation.  Combining the best human (Expert 3) with the best LLM (Gemini 2.5 Pro) yields $\rho = 0.67$, and this null result holds across all human-LLM pairings we examined.  However, our design tests only ex post combination of independent judgments; real-time collaboration, where humans and AI interact during the prediction process, might yield different results and remains an important direction for future research.

In summary, aggregation does not improve upon the best individual model in our sample.  The only pairwise combination that marginally improves performance (Gemini 2.5 Pro plus GPT-5 Mini) reflects modest complementarity between two already-similar high performers.  In general, aggregating strong evaluators with weaker ones degrades performance, suggesting that selecting the single best model dominates ensemble strategies for this task.

\section{Robustness Sample and Task Difficulty}\label{app:robustness}

To assess the robustness of our findings, we constructed a second independent sample of Kickstarter projects using the same selection procedure as for the initial sample.\footnote{Preregistration of this subsequent study can be found here: https://zenodo.org/records/17703676} Four projects were subsequently excluded because they were removed from the platform, leaving 26 projects in the robustness sample compared to 30 in the initial sample.
Table~\ref{tbl:kickstarter_projects_robust} includes the list of projects included in the robustness analysis along with descriptive information.

\begin{sidewaystable}
\centering
\scriptsize
\begin{tabularx}{\textwidth}{@{}llllrrX@{}}
\toprule
\textbf{ID}    & \textbf{Subcategory} & \textbf{Deadline} & \textbf{Launch} & \textbf{Goal} & \textbf{Raised} & \textbf{URL} \\
\midrule
bunker & Apps & 11/26/2025 & 10/27/2025 & \$23,700 & \$5,145 & \href{https://www.kickstarter.com/projects/bunkerbros/the-bunker-app}{\nolinkurl{https://www.kickstarter.com/projects/bunkerbros/the-bunker-app}} \\
adoptee & Software & 11/26/2025 & 11/01/2025 & \$25,000 & \$26,000 & \href{https://www.kickstarter.com/projects/378078478/adoptee-identity}{\nolinkurl{https://www.kickstarter.com/projects/378078478/adoptee-identity}} \\
techbee & Hardware & 11/27/2025 & 11/12/2025 & \$10,000 & \$84,064 & \href{https://www.kickstarter.com/projects/1309949597/techbee-smoker}{\nolinkurl{https://www.kickstarter.com/projects/1309949597/techbee-smoker}} \\
topnyne & Apps & 11/28/2025 & 10/29/2025 & \$30,000 & \$255 & \href{https://www.kickstarter.com/projects/topnyne/topnyne-the-media-app-for-entertainment-and-monetization}{\nolinkurl{https://www.kickstarter.com/projects/topnyne/topnyne-the-media-app-for-entertainment-and-monetization}} \\
neuro & Wearables & 11/30/2025 & 10/26/2025 & \$18,000 & \$21,338 & \href{https://www.kickstarter.com/projects/generalneuro/neurolingo-neurotech-for-your-second-language}{\nolinkurl{https://www.kickstarter.com/projects/generalneuro/neurolingo-neurotech-for-your-second-language}} \\
caira & Camera Equiptment & 12/02/2025 & 11/04/2025 & \$20,000 & \$459,668 & \href{https://www.kickstarter.com/projects/cameraintelligence/caira-worlds-first-ai-native-mirrorless-camera}{\nolinkurl{https://www.kickstarter.com/projects/cameraintelligence/caira-worlds-first-ai-native-mirrorless-camera}} \\
spyfinder & Gadgets & 12/03/2025 & 11/03/2025 & \$25,000 & \$2,740 & \href{https://www.kickstarter.com/projects/spyassociates/spyfinder-proscan-bluetooth-gps-tag-rf-detector}{\nolinkurl{https://www.kickstarter.com/projects/spyassociates/spyfinder-proscan-bluetooth-gps-tag-rf-detector}} \\
elejoy & Hardware & 12/04/2025 & 11/04/2025 & \$20,000 & \$67,780 & \href{https://www.kickstarter.com/projects/elejoy-v2h-inverter/elejoy-v2h-turn-your-tesla-into-your-home-backup-powerhouse}{\nolinkurl{https://www.kickstarter.com/projects/elejoy-v2h-inverter/elejoy-v2h-turn-your-tesla-into-your-home-backup-powerhouse}} \\
mask & Wearables & 12/04/2025 & 10/05/2025 & \$30,000 & \$45 & \href{https://www.kickstarter.com/projects/kbeautycos/next-gen-k-beauty-acne-mask}{\nolinkurl{https://www.kickstarter.com/projects/kbeautycos/next-gen-k-beauty-acne-mask}} \\
neotron & Sound & 12/05/2025 & 11/10/2025 & \$400,000 & \$300 & \href{https://www.kickstarter.com/projects/allonesun/neotron-the-touchscreen-electronic-music-groove-box}{\nolinkurl{https://www.kickstarter.com/projects/allonesun/neotron-the-touchscreen-electronic-music-groove-box}} \\
kara & Hardware & 12/06/2025 & 10/21/2025 & \$50,000 & \$414,222 & \href{https://www.kickstarter.com/projects/karawater/karapure-2-air-to-alkaline-water-dispenser-w-ultrafiltration}{\nolinkurl{https://www.kickstarter.com/projects/karawater/karapure-2-air-to-alkaline-water-dispenser-w-ultrafiltration}} \\
reader & Apps & 12/06/2025 & 11/06/2025 & \$99,000 & \$10 & \href{https://www.kickstarter.com/projects/radiant-epistemology/radiant-reader-lighting-up-the-future-of-reading}{\nolinkurl{https://www.kickstarter.com/projects/radiant-epistemology/radiant-reader-lighting-up-the-future-of-reading}} \\
kicks & Apps & 12/07/2025 & 11/07/2025 & \$80,000 & \$40 & \href{https://www.kickstarter.com/projects/kicksstand/kicks-standcom-the-ultimate-sneakerhead-social-platform}{\nolinkurl{https://www.kickstarter.com/projects/kicksstand/kicks-standcom-the-ultimate-sneakerhead-social-platform}} \\
roofpilot & Web & 12/08/2025 & 10/30/2025 & \$20,000 & \$0 & \href{https://www.kickstarter.com/projects/roofpilot/roofpilot-launch-manage-or-grow-a-roofing-company}{\nolinkurl{https://www.kickstarter.com/projects/roofpilot/roofpilot-launch-manage-or-grow-a-roofing-company}} \\
lightbar & DIY Electronics & 12/11/2025 & 10/27/2025 & \$250,485 & \$4,003 & \href{https://www.kickstarter.com/projects/mooolink/lightbar}{\nolinkurl{https://www.kickstarter.com/projects/mooolink/lightbar}} \\
brushly & Software & 12/11/2025 & 11/11/2025 & \$20,000 & \$101 & \href{https://www.kickstarter.com/projects/brushly/brushlyart-your-co-pilot-for-a-full-time-art-career}{\nolinkurl{https://www.kickstarter.com/projects/brushly/brushlyart-your-co-pilot-for-a-full-time-art-career}} \\
refood & Apps & 12/11/2025 & 10/19/2025 & \$10,000 & \$11 & \href{https://www.kickstarter.com/projects/refood2025/refood-the-smart-food-connection-app}{\nolinkurl{https://www.kickstarter.com/projects/refood2025/refood-the-smart-food-connection-app}} \\
6-pound & Gadgets & 12/12/2025 & 10/13/2025 & \$75,000 & \$26,113 & \href{https://www.kickstarter.com/projects/458223518/6-pound-phone-case}{\nolinkurl{https://www.kickstarter.com/projects/458223518/6-pound-phone-case}} \\
chalkdrop & Hardware & 12/12/2025 & 11/12/2025 & \$10,000 & \$67 & \href{https://www.kickstarter.com/projects/172280277/chalkdrop-the-tiny-machine-that-eats-carbon}{\nolinkurl{https://www.kickstarter.com/projects/172280277/chalkdrop-the-tiny-machine-that-eats-carbon}} \\
bantxr & Apps & 12/13/2025 & 10/29/2025 & \$90,000 & \$122 & \href{https://www.kickstarter.com/projects/bantxr/make-bantxr-happen}{\nolinkurl{https://www.kickstarter.com/projects/bantxr/make-bantxr-happen}} \\
cooler & Gadgets & 12/14/2025 & 11/14/2025 & \$12,450 & \$2,005 & \href{https://www.kickstarter.com/projects/williamwarwickiv/the-ghost-lifetime-cooler-patented-space-age-insulation}{\nolinkurl{https://www.kickstarter.com/projects/williamwarwickiv/the-ghost-lifetime-cooler-patented-space-age-insulation}} \\
heavyfoot & Hardware & 12/15/2025 & 11/05/2025 & \$96,700 & \$200 & \href{https://www.kickstarter.com/projects/hfw-2019/heavyfoot-weights-more-fun-less-worry}{\nolinkurl{https://www.kickstarter.com/projects/hfw-2019/heavyfoot-weights-more-fun-less-worry}} \\
where-app & Apps & 12/15/2025 & 11/15/2025 & \$50,000 & \$273 & \href{https://www.kickstarter.com/projects/whereapp-ron/whereapp-experience-the-world-share-it}{\nolinkurl{https://www.kickstarter.com/projects/whereapp-ron/whereapp-experience-the-world-share-it}} \\
godogolem & Software & 12/17/2025 & 11/02/2025 & \$55,000 & \$60 & \href{https://www.kickstarter.com/projects/godogolem/godogolem}{\nolinkurl{https://www.kickstarter.com/projects/godogolem/godogolem}} \\
eclep & Apps & 12/17/2025 & 11/17/2025 & \$116,000 & \$0 & \href{https://www.kickstarter.com/projects/eclep-founder/eclep-ai-powered-mind-and-body-app-for-conscious-living}{\nolinkurl{https://www.kickstarter.com/projects/eclep-founder/eclep-ai-powered-mind-and-body-app-for-conscious-living}} \\
hydro & 3D Printing & 12/18/2025 & 10/25/2025 & \$25,000 & \$372 & \href{https://www.kickstarter.com/projects/turinginnovations/sectional-modular-vertical-hydroponic-farm-patent-pending}{\nolinkurl{https://www.kickstarter.com/projects/turinginnovations/sectional-modular-vertical-hydroponic-farm-patent-pending}} \\
\bottomrule
\end{tabularx}

\caption{Kickstarter projects included in the robustness sample, with funding goals, amounts raised, campaign dates, and URLs.}
\label{tbl:kickstarter_projects_robust}
\end{sidewaystable}

\begin{figure}[tb]
\centering
\includegraphics[width=\linewidth]{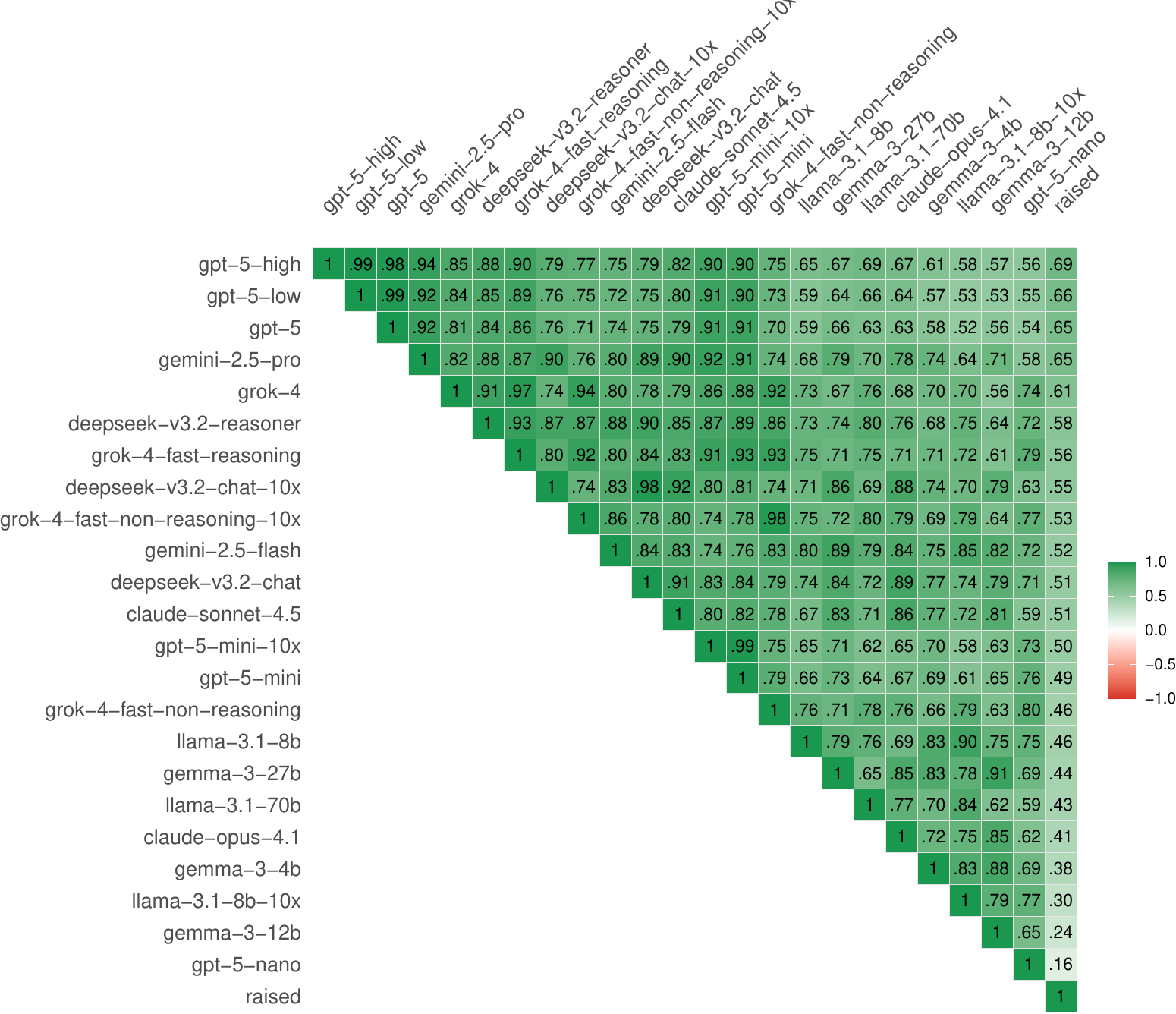}
\caption{Spearman correlation matrix among all evaluators in the robustness sample.  The rightmost column shows each evaluator's correlation with actual funds raised.}\label{fig:corr_table_robust}
\end{figure}

Figure~\ref{fig:corr_table_robust} reports Spearman rank correlations between each evaluator's predicted rankings and realized fundraising outcomes for the robustness sample.  At a high level, we see that performance is slightly lower overall, while the relative ordering of model performance remains similar across samples.

\subsection{Task Difficulty and Outcome Compression}

Differences in absolute performance across samples must be interpreted in light of task difficulty.  For rank-based evaluation, difficulty depends not only on predictor quality but also on how distinguishable the realized outcomes are from one another.  When many outcomes are closely clustered, even small perturbations in predictions can generate rank reversals, mechanically reducing achievable rank correlations.

The robustness sample exhibits substantially greater outcome compression at the lower end of the distribution.  In the robustness sample, 8 of 26 projects (30.8\%) raised less than \$100, compared to 6 of 30 projects (20.0\%) in the initial sample.  The robustness sample also includes two projects that raised \$0 and several additional projects clustered at \$10--\$11, whereas the initial sample contains no zero-raised projects.

This compression translates directly into greater pairwise ambiguity.  In the initial sample, only 23 of 435 project pairs (5.3\%) differ in realized funds raised by \$100 or less.  In contrast, 49 of 325 pairs (15.1\%) in the robustness sample fall within this range.  Thus, the robustness sample contains nearly three times as many near-indistinguishable project pairs, making accurate ordinal prediction intrinsically more difficult.

\subsection{Performance Stability Across Samples}

Consistent with this increase in difficulty, absolute Spearman correlations are generally lower in the robustness sample.  However, the relative ordering of model performance remains moderately stable.  Across the 19 models appearing in both samples, the Spearman rank correlation between model performance rankings in the initial and robustness samples is approximately 0.54.  Models that perform best in the initial sample---including the GPT-5 variants and Gemini~2.5~Pro---remain among the strongest performers in the robustness sample, despite lower absolute correlations.

Taken together, these results indicate that raw performance metrics are not directly comparable across samples with different outcome distributions.  Lower absolute correlations in the robustness sample reflect, in part, a more compressed and ambiguous outcome space rather than a fundamental breakdown of predictive capability.  The persistence of relative performance rankings across samples therefore provides additional support for the robustness of our main conclusions.

\section{Aggregating Multiple Runs}\label{app:large_n}

A natural question is whether running a less expensive model multiple times and aggregating its predictions can substitute for a single run of a more capable model.  To explore this possibility, we selected four models and ran each model ten times on the robustness sample, summing wins across the tournaments.\footnote{Preregistration of this subsequent study can be found here: https://zenodo.org/records/17704106}

\begin{figure}[!tb]
\centering
\includegraphics[width=0.8\linewidth]{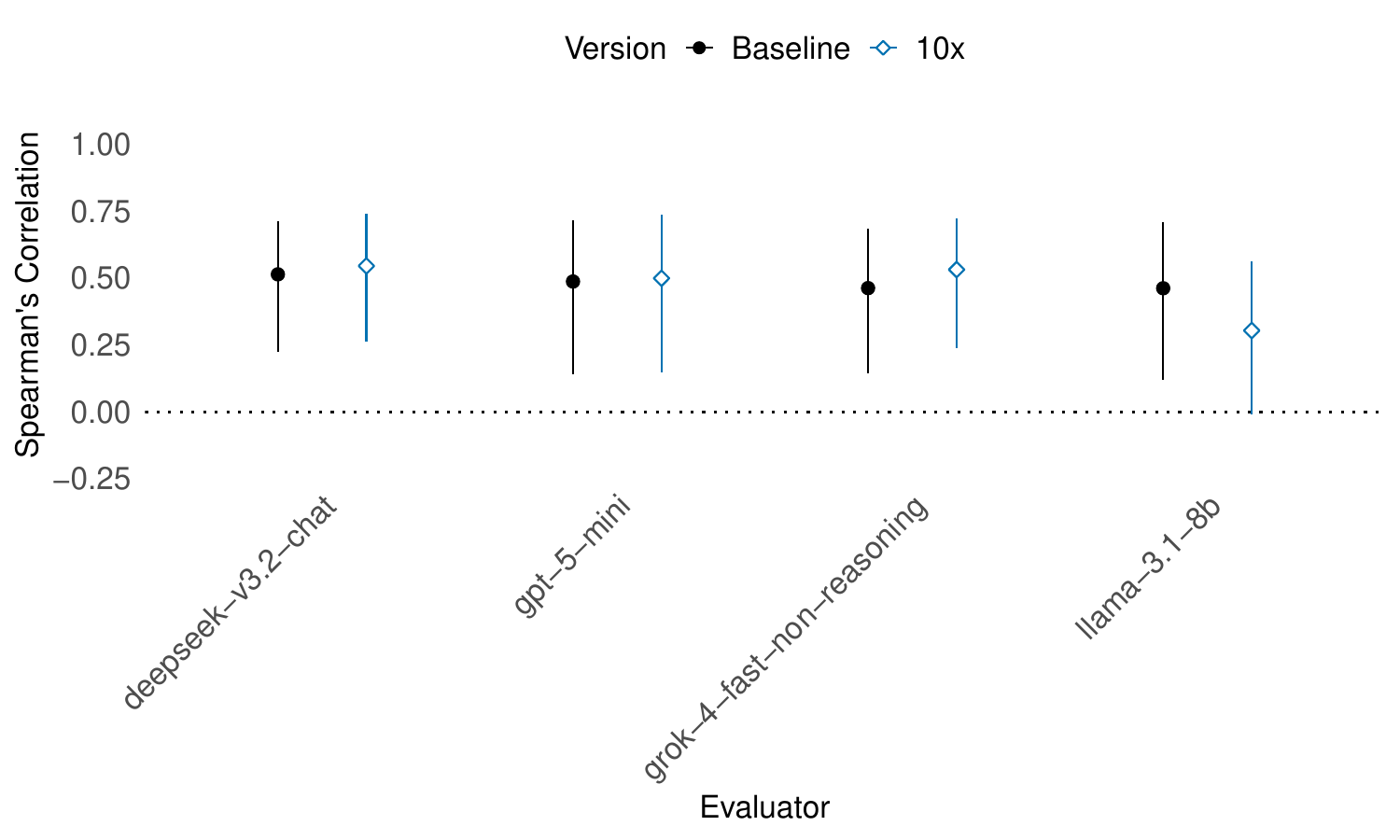}
\caption{Spearman's rank correlation ($\rho$) between predicted and realized project rankings for single runs (black dots) and ten-run aggregates (blue diamonds), with 90\% confidence intervals.  The dashed line at zero represents chance performance.}
\label{fig:large_n}
\end{figure}

Figure~\ref{fig:large_n} reports Spearman's rank correlation ($\rho$) between predicted and realized project rankings for single-run evaluations and for the corresponding ten-run aggregated predictions.  Across the four models tested, aggregation yields mixed and modest effects.  Three models exhibit small improvements in $\rho$ under aggregation (DeepSeek-v3.2-chat: 0.51 to 0.55; GPT-5-mini: 0.49 to 0.51; Grok-4-fast-non-reasoning: 0.46 to 0.53), while one model (Llama-3.1-8b) experiences a more substantial decline (0.46 to 0.30).  Averaged across models, the change in $\rho$ is effectively zero ($\Delta\rho \approx -0.01$), and all changes fall within the corresponding bootstrap confidence intervals.

Crucially, even the strongest aggregated result ($\rho \approx 0.55$) remains well below the performance of top frontier models evaluated with a single run, such as GPT-5-high ($\rho \approx 0.69$) and Gemini~2.5~Pro ($\rho \approx 0.63$) on the same sample.  Thus, repeated runs of weaker models do not close the performance gap with more capable models.

The limited and inconsistent gains from aggregation are consistent with the statistical requirement that aggregation benefits depend on the independence of errors.  Repeated runs of the same model are highly correlated: they share architecture, training data, and prompt interpretation, differing primarily through stochastic token sampling.  For $n$ observations with pairwise correlation $\rho$, the effective sample size is $n_{\text{eff}} = n / \bigl(1 + (n-1)\rho\bigr)$.  If inter-run correlations are high---as is plausible for repeated runs of the same model---aggregation yields little additional information.  For example, at $\rho = 0.9$, ten runs correspond to an effective sample size of approximately 1.1.

From this perspective, prediction error can be decomposed into systematic bias, which is stable across runs, and idiosyncratic noise, which varies with sampling.  Aggregation can reduce only the latter.  The near-zero average improvement observed here suggests that systematic model limitations dominate performance in this task.  For practitioners, these results imply that resources may be better allocated to more capable models than to repeated runs of less capable ones.

\end{document}